\documentclass[12pt]{article}
\usepackage{titling}
\usepackage{amsmath}
\usepackage{slashed}
\usepackage{amssymb}
\usepackage{epsfig}
\usepackage{graphicx}
\usepackage{multirow}
\usepackage{color}
\usepackage{subcaption}
\usepackage{rotating}
\usepackage{hyperref}
\usepackage[margin=1.0in]{geometry}
\usepackage[table]{xcolor}
\usepackage{enumitem}
\usepackage[utf8x]{inputenc}
\usepackage[compress,numbers,sort]{natbib}
\usepackage{authblk}
\usepackage{colortbl}
\usepackage{pdflscape}
\usepackage{color}
\usepackage{float}
\usepackage{siunitx}
\usepackage[compat=1.1.0]{tikz-feynman}
\usepackage{lscape}
\usepackage{mathtools}
\usepackage{amsfonts}
\usepackage{bbold}
\usepackage{soul}
\usepackage{comment}

\definecolor{nicered}{rgb}{0.6,0.1,0.1}
\definecolor{nicegreen}{rgb}{0.1,0.5,0.1}
\definecolor{mediumcandyapplered}{rgb}{0.99, 0.12, 0.07}
\definecolor{red}{rgb}{1.0, 0, 0}
\hypersetup{colorlinks,citecolor= nicegreen,linkcolor= nicered}

\renewcommand{\bar}{\overline}

\definecolor{LightCyan}{rgb}{0.88,1,1}
\definecolor{piggypink}{rgb}{0.99, 0.87, 0.9}
\definecolor{applegreen}{rgb}{0.55, 0.71, 0.0}
\definecolor{darkpastelgreen}{rgb}{0.01, 0.75, 0.24}
\definecolor{green-yellow}{rgb}{0.68, 1.0, 0.18}

\newcommand{\beq}{\begin{equation}}
	\newcommand{\eeq}{\end{equation}}
\newcommand{\bea}{\begin{eqnarray}}
	\newcommand{\eea}{\end{eqnarray}}

\DeclareSIUnit\barn{b}
\newcommand{\cC}{\mathcal{C}}
\newcommand{\cO}{\mathcal{O}}
\newcommand{\cL}{\mathcal{L}}

\newcommand{\hc}{\text{h.c.}}
\newcommand{\lsquare}{\left[}
\newcommand{\rsquare}{\right]}
\newcommand{\lround}{\left(}
\newcommand{\rround}{\right)}
\newcommand{\lambdaNPNP}[2]{\lambda_{#1 #2}}

\title{\bf{Probing New Physics with the Electron Yukawa coupling} 
}

\author[1,2]
{Barbara Anna Erdelyi \thanks{barbaraanna.erdelyi@pd.infn.it}}
\author[1,2]
{Ramona Gr\"ober \thanks{ramona.groeber@pd.infn.it}}
\author[2]
{Nud\v zeim Selimovi\'c \thanks{nudzeim.selimovic@pd.infn.it}}

\affil[1]{\emph{\normalsize Dipartimento di Fisica e Astronomia ``G. Galilei'', Universit\`a di Padova, Padova, Italy}}
\affil[2]{\emph{\normalsize Istituto Nazionale di Fisica Nucleare, Sezione di Padova, Padova, Italy}}

\date{}

\begin{document}
	\maketitle
\begin{abstract}
	\normalsize
    A dedicated run of a future electron-positron collider (FCC-ee) at a center-of-mass energy equal to the Higgs boson mass would enable a direct measurement of the electron Yukawa coupling. However, it poses substantial experimental difficulties due to large backgrounds, the requirement for monochromatised $e^+e^-$ beams, and the potential extension of the FCC-ee timeline. Given this, we explore the extent to which the electron Yukawa coupling can be enhanced in simplified UV models and examine whether such scenarios can be constrained by other FCC-ee runs or upcoming experiments at the intensity frontier. Our results indicate that in certain classes of models, the $(g-2)_e$ provides a probe of the electron Yukawa coupling that is as effective or better than the FCC-ee. Nevertheless, there exist models that can lead to sizeable deviations in the electron Yukawa coupling which can only be probed in a dedicated run at the Higgs pole mass.
\end{abstract}	
	
\clearpage
{
	\hypersetup{linkcolor=black}
	\tableofcontents
}
	
\section{Introduction} 
\label{sec:introduction}

The Higgs boson, as the cornerstone of the Standard Model (SM) of particle physics, offers a unique window into the mechanisms of electroweak symmetry breaking and the origin of particle masses. While many of its properties have been measured with remarkable precision \cite{ATLAS:2022vkf,CMS:2022dwd}, its smallest coupling in the SM, the coupling to electrons, remains largely unexplored. Searches for Higgs decays to $e^+e^-$ constrained the branching ratio (BR) of the Higgs boson to 
electrons to $\text{BR}(h \to e^{+}e^{-}) < 3.6 \cdot 10^{-4}$, implying an upper limit on the electron Yukawa of 260 times its SM value \cite{ATLAS:2019old}. At the HL-LHC, the projected sensitivity for the electron Yukawa coupling is $y_e< 120 \,y_e^{\rm SM}$ \cite{Cepeda:2019klc}. Indeed, it turns out that a measurement of the SM electron Yukawa coupling via the Higgs boson decay $h\to e^+ e^-$ is challenging due to the dominance of the Dalitz decays $h \to e^+ e^-\gamma$ \cite{Sun:2013rqa}. Furthermore, the product of nucleon and electron Yukawa couplings can be constrained in atomic spectroscopy \cite{Delaunay:2016brc}, providing though only very weak limits.
\par
Given this, the only possible strategy to measure the electron Yukawa coupling seems to be a dedicated run via $s$ channel production of a Higgs boson at a future $e^+e^-$ collider with center-of-mass energy equal to the Higgs boson mass. Such a measurement is highly challenging, due to large backgrounds, the requirement that the $e^+e^-$ beams must be monochromatized, and the need for a precise knowledge of the Higgs boson mass to hit the resonance. With two years of runtime of the FCC-ee at this energy and by combining four interaction points, a bound of $|y_e|<1.6\, y_e^{\rm SM}$ at 95\% confidence level (CL) could be reached~\cite{dEnterria:2021xij}. In Ref.~\cite{Boughezal:2024yjk} it was shown that the significances could be potentially increased with polarised beams in measurements of single transverse-spin asymmetries.
\par 
Considering the challenges and extended timeline associated with measuring the electron Yukawa coupling, this work seeks to address the question of whether pursuing such an endeavor is worthwhile. To this end, we investigate models in which deviations in the electron coupling originate from dimension-six operators in the Standard Model Effective Field Theory (SMEFT) \cite{Brivio:2017vri, Grzadkowski:2010es}.\footnote{We emphasise that more moderate deviations in the electron Yukawa coupling can be achieved by radiative corrections to the electron Yukawa coupling or by effective operators of higher dimension.} Following the approach outlined in \cite{Erdelyi:2024sls}, we identify three classes of ultraviolet (UV) models based on the spin of new states that yield deviations in the electron Yukawa coupling which are chirally enhanced by a large factor of $\mathcal{O}(v/m_e)$. Some of the models we will consider here have already been studied in the literature in the context of enhanced electron Yukawa couplings, in particular, the two-Higgs doublet model (2HDM) in Ref.~\cite{Dery:2017axi} and a model with an extra vector-like lepton (VLL) and a scalar singlet in Ref.~\cite{Davoudiasl:2023huk}.

The considered new physics (NP) scenarios produce additional effective operators constrained by Higgs physics, electroweak precision data, flavour physics, and direct searches. We specifically investigate whether such scenarios can be adequately constrained by other upcoming experimental efforts at both high-intensity and high-energy scales, including future runs at the FCC-ee. As has been shown in \cite{Erdelyi:2024sls,Allwicher:2024sso,Greljo:2024ytg,Gargalionis:2024jaw} for simplified models and in \cite{DeBlas:2019qco,Celada:2024mcf,Maura:2024zxz} for EFT parameterisations, the extreme precision of the FCC-ee run at the $Z$-pole mass can probe new physics scales well above the TeV scale. We show that the interplay with high-intensity probes, in particular the $(g-2)$ of the electron, can in some cases be stronger, or as strong, as the FCC-ee constraints.  

This study is organised as follows: In Sec.~\ref{sec:effectivefieldtheory&UVmodels} we identify three classes of possible models that can enhance the electron Yukawa coupling. Then, in Sec.~\ref{sec:constraints} we discuss how direct searches together with the electroweak, flavour, and Higgs physics observables constrain the simplified models that we consider. The derived bounds are then applied in Sec.~\ref{sec:results}. Finally, we present our conclusions in Sec.~\ref{sec:conclusions}.

% % % % % % % % % % % % 
\section{Classification of the Simplified Models}
\label{sec:effectivefieldtheory&UVmodels}
We start by identifying the minimal extensions of the SM which can cause an enhancement of the effective coupling between the electron and the Higgs boson. We assume a separation of scales between the masses of the new particles and the electroweak scale, motivating the use of the SMEFT to describe the effects of new physics.  The SMEFT Lagrangian, $\cL_{\rm SMEFT}$, is constructed by adding to the SM Lagrangian, $\cL_{\rm SM}$, all higher-dimensional operators consistent with the symmetries of the SM. The Lagrangian reads
\begin{equation}
    \mathcal{L}_{\rm SMEFT}=\mathcal{L}_{\rm SM}+\sum_{d=5}^{\infty} \cC_k^{(d)} \cO_k^{(d)}\,, \quad \text{where} \, \lsquare \cC_k^{(d)}\rsquare = 4-d\,.
\end{equation}
Notationwise, the symbol $d$ denotes the operator's ($\cO_k^{(d)}$) mass dimension,  $k$ runs over all operators at fixed dimension $d$, and  $\cC^{(d)}_k$ are the Wilson coefficients associated with each operator. Additionally, we adopt the following notation for the SM Lagrangian
\begin{align}\label{eq:LagSM}
\mathcal{L}_{\rm SM} &=\, -\frac{1}{4}B_{\mu\nu}B^{\mu\nu} - \frac{1}{4}W_{\mu\nu}^I W^{I\mu\nu} - \frac{1}{4}G_{\mu\nu}^A G^{A\mu\nu}+D_\mu \phi^\dag D^\mu \phi -V(\phi)
\nonumber\\
&\phantom{aai}+\sum_{\psi} \bar\psi i\slashed{D}\psi-\left[y_d \,\bar{q}_L \phi\, d_R+ y_u\, \bar q_L  \tilde \phi\,  u_R +  y_e\,\bar  \ell_L \phi \,e_R+\text{h.c.}\right]\,,\\
V(\phi) &= -\mu^2 \phi^\dag \phi +\lambda (\phi^\dag \phi)^2\,,
\end{align}
where the field strength tensors $G_{\mu\nu}$, $W_{\mu\nu}$, and $B_{\mu\nu}$ are associated respectively with the SM gauge group $G_{\rm SM}=SU(3)\times SU(2)\times U(1)$. The sum over $\psi$ runs over the chiral SM fermions: quark and lepton $SU(2)$ doublets $q_L$, $\ell_L$, and the singlets $u_R$, $d_R$, $e_R$. The Yukawa sector of the SM Lagrangian describes interactions between fermions and the $SU(2)$ Higgs doublet $\phi$ ($\tilde{\phi}=i\sigma_2 \phi^*$) and the associated Yukawa couplings are $3\times3$ complex matrices in flavour space $y_i$ (with $i=u,d,e$). Upon electroweak symmetry breaking, the Higgs boson acquires a vacuum expectation value (vev) $v=|\mu|/\sqrt{\lambda}=246$ GeV. 

In the SMEFT framework, the direct relationship between the electron Yukawa coupling and its mass, as predicted by the SM, is disrupted by higher-dimensional operators. In particular, the operator 
\begin{equation}
    \cO_{e\phi}= \left(\phi^\dagger \phi\right) \left(\bar{l}_L \phi \,e_R\right)\,,
    \label{eq:dimensionsix_Oephi}
\end{equation}
is the leading contribution to the effective lepton masses and the Yukawa couplings
\begin{align}
    [m_{e}^{\rm eff}]_{ij} &= \frac{v}{\sqrt{2}} \left([y_e]_{ij} - \frac{v^2}{2} [\mathcal{C}_{e\phi}]_{ij}\right)
    \label{eq:eff_el_mass}\,,\\
    [y_{e}^{\rm eff}]_{ij} & = \frac{1}{\sqrt{2}}\left([y_e]_{ij}-\frac{3v^2}{2}[\mathcal{C}_{e\phi}]_{ij}\right)
    \label{eq:eff_el_Yuk}\,,
\end{align}
where $[\mathcal{C}_{e\phi}]_{ij}$ is the associated Wilson coefficient. Upon rotating to the mass eigenbasis, the electron Yukawa coupling in SMEFT up to dimension six is expressed as
\begin{equation}
    g_{hee}= \frac{m_e}{v}\left[1+ v^2 \cC_{\phi,{\rm kin}}\right]-\frac{v^2}{\sqrt{2}} {[\tilde{\cC}_{e\phi}]_{11}}\,,
    \label{eq:ghee}
\end{equation}
where $\tilde{\cC}_{e\phi}$ is obtained from $\cC_{e\phi}$ by the rotation to the mass basis. The term $\cC_{\phi,{\rm kin}}$ reads
\begin{equation}
    \cC_{\phi,{\rm kin}}=\left(\cC_{\phi\Box}-\frac{1}{4}C_{\phi D} \right)\,,
    \label{eq:C_kin}
\end{equation}
and captures contributions from derivative Higgs operators
\begin{equation}
    \cO_{\phi\Box} = (\phi^{\dagger}\phi) \Box (\phi^{\dagger}\phi)\,,\quad \quad \cO_{\phi D}=|\phi^{\dagger} D_{\mu} \phi|^2\,,
    \label{eq:dimensionsix_D2phi4}
\end{equation}
which modify the Higgs kinetic term and influence all Higgs processes after the appropriate field redefinition. However, unlike $\tilde{\cC}_{e\phi}$, these contributions always retain the $m_e/v \simeq 10^{-6}$ suppression intrinsic to the SM electron Yukawa coupling. In addition, since they modify all the Higgs couplings, they can be best constrained when measuring the couplings of the Higgs boson to heavier particles.

To better quantify the effects of the heavy new physics, we define the electron Yukawa modifier as
\begin{equation}
    \kappa_{e} = \frac{g_{hee} v}{m_e} = 1 - \frac{v^3}{\sqrt{2}\, m_{e}} \lsquare \tilde{\cC}_{e\phi} \rsquare_{11}\,.
    \label{eq:kappa_def}
\end{equation}
Interestingly, new physics at the TeV scale with $\mathcal{O}(1)$ couplings to SM fields could yield a correction of $\delta g_{hee} \simeq v^2/\Lambda^2 \simeq 4 \cdot 10^{-2}$, leading to $\kappa_e \simeq 2 \cdot 10^{5}$. However, new physics interacting with electrons is typically subject to stringent constraints from various processes. This work aims to precisely determine the potential value of $\kappa_e$ in simplified new physics scenarios that could significantly enhance it.

Before delving into the identification of such scenarios, it is important to note that models predicting large values of $\kappa_e$ inherently require a degree of fine-tuning, depending on the size of $\kappa_e$. From Eqs.~\eqref{eq:eff_el_mass} and~\eqref{eq:eff_el_Yuk}, it becomes evident that maintaining the experimentally observed small electron mass necessitates a cancellation between the two terms in Eq.~\eqref{eq:eff_el_mass}, dictated by the size of $y_{e}^{\rm eff}$ in Eq.~\eqref{eq:eff_el_Yuk}. Furthermore, for the same reason, the renormalizable Yukawa coupling $y_e$ is expected to approximate $v^2 \cC_{e\phi}/2$, aligning its size with that of the dimension-six new physics contributions within the EFT framework.

Now we turn to identify simplified models that lead to tree-level contributions to $\mathcal{C}_{e\phi}$, beginning with single-field extensions of the SM. However, aside from the case of the new scalar $SU(2)$ doublet listed in Tab.~\ref{tab:newphysicsscalars}, all single field extensions that match onto $\cO_{e\phi}$ at tree level generate the corresponding coefficient proportional to the renormalisable Yukawa coupling 
\begin{equation}
    [\cC_{e\phi}]_{ij} \simeq [y_e]_{ij} \frac{\lambda_{\rm NP}^2}{\Lambda^2}\,,
\end{equation}
where $\lambda_{\rm NP}$ serves as a placeholder representing the couplings of NP to SM states. Consequently, this implies that the effective Yukawa coupling lacks chiral enhancement, and any contribution to $\kappa_e$ will be limited to
\begin{equation}
    \kappa_e \simeq 1 \pm v^2 \frac{\lambda_{\rm NP}^2}{\Lambda^2}\,,
\end{equation}
where weakly coupled new physics at the TeV scale leads to corrections below 10\%, rendering such effects unobservable experimentally. Moreover, in all concrete scenarios involving vector-like leptons, as detailed in Tab.~\ref{tab:vectorlikeleptons}, the contribution to $[\cC_{e\phi}]_{ij}$ is inherently linked to the contributions to one of the following operators
\begin{align}
    \lsquare\cO_{\phi\ell}^{(1)}\rsquare_{ij} & = (i\phi^\dagger \overleftrightarrow{D}_\mu \phi)(\bar{\ell}_L^i \gamma^\mu \ell_L^j)\,,\label{eq:Ophil1}\\
    \lsquare\cO_{\phi\ell}^{(3)}\rsquare_{ij} &= (i\phi^\dagger \overleftrightarrow{D}_\mu^I \phi)(\bar{\ell}_L^i \gamma^\mu \sigma^I \ell_L^j) \,,\label{eq:Ophil3}\\
    \lsquare\cO_{\phi e}^{\phantom{(1)}}\rsquare_{ij} & = (i\phi^\dagger \overleftrightarrow{D}_\mu \phi)(\bar{e}_R^i \gamma^\mu e_R^j)\,,\label{eq:Ophie}
\end{align}
which modify $Z$ interactions to leptons and are strongly constrained. To illustrate this, let us consider the scenario where only the heavy partner of the right-handed electron, $E\sim (\mathbf{1},\mathbf{1})_{-1}$, is introduced. In this case, the contribution to $\kappa_e$ is expressed as
\begin{equation}
    \kappa_e = 1 - 2\delta g_{L11}^{Ze} \,,
\end{equation}
where $\delta g_{L11}^{Ze}$ represents the modification of the $Z$-coupling to electrons induced by this state and defined in Eq.~\eqref{eq:deltagZeL}. This modification is constrained to the per-mille level, making the prospect of detecting such a deviation in $\kappa_e$ poor.  A similar outcome arises for all other single mediator extensions, with the exception of $\varphi$ which will be discussed below.

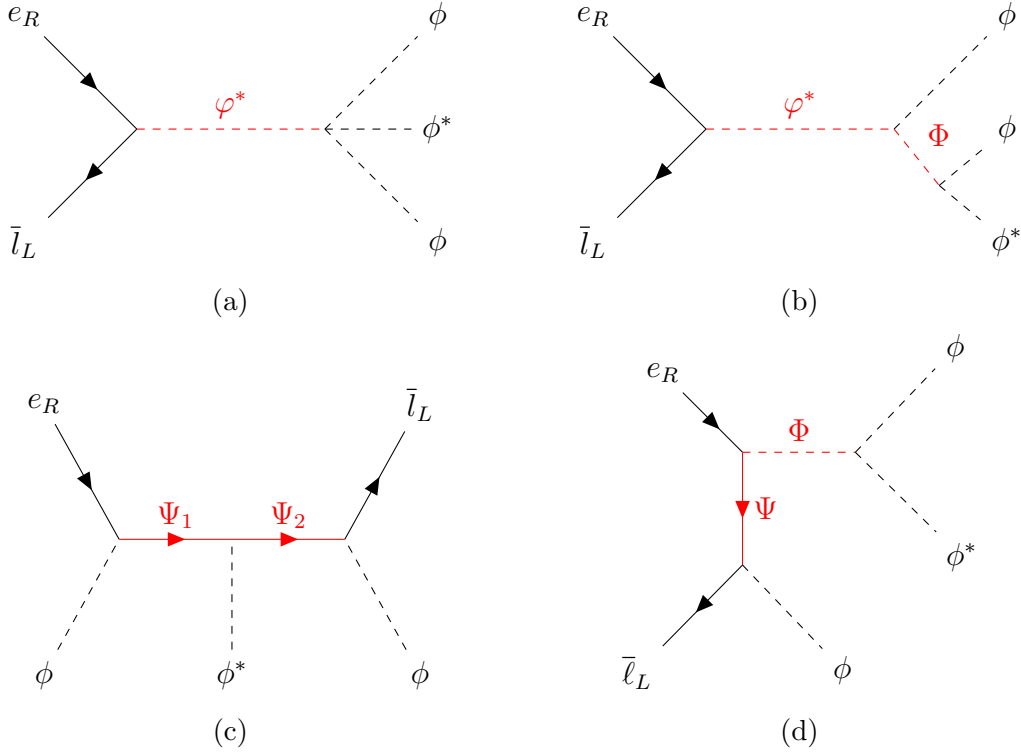
\begin{figure}[t]
	\centering
	\begin{subfigure}[t]{0.45\linewidth}\centering
		\begin{tikzpicture}
			\begin{feynman}
				\vertex (a) at (0,0);
				\vertex(b) at (2.5,0);
				\vertex (lL) at (-1.5,-1.5) {\(\bar{l}_L\)};
				\vertex (eR) at (-1.5,1.5) {\(e_R\)};
				\vertex (phi1) at (4,1.5) {\(\phi\)};
				\vertex (phi2) at (4,0) {\(\phi^*\)};
				\vertex (phi3) at (4,-1.5) {\(\phi\)};
				\diagram*{
					(eR) -- [fermion] (a) -- [fermion] (lL),
					(a) -- [scalar, red, edge label = \(\varphi^*\)] (b);
					(b) -- [scalar] (phi1);
					(b) -- [scalar] (phi2);
					(b) -- [scalar] (phi3);
				};
			\end{feynman}
		\end{tikzpicture}
		\caption{}\label{subfig:UVmodelssecondHiggs}
	\end{subfigure}
	\begin{subfigure}[t]{0.45\linewidth}\centering
		\begin{tikzpicture}
			\begin{feynman}
				\vertex (a) at (0,0);
				\vertex (b) at (2.5,0);
				\vertex (lL) at (-1.5,-1.5) {\(\bar{l}_L\)};
				\vertex (eR) at (-1.5,1.5) {\(e_R\)};
				\vertex (phi1) at (4,1.5) {\(\phi\)};
				\vertex (S) at (3.1,-0.75);
				\vertex (phi2) at (4,-0.) {\(\phi\)};
				\vertex (phi3) at (4,-1.5) {\(\phi^*\)};
				\diagram*{
					(eR) -- [fermion] (a) -- [fermion] (lL),
					(a) -- [scalar, red, edge label = \(\varphi^*\)] (b);
					(b) -- [scalar] (phi1);
					(b) -- [scalar, red, edge label= \(\Phi\)] (S);
					(S) -- [scalar] (phi2);
					(S) -- [scalar] (phi3);
				};
			\end{feynman}
		\end{tikzpicture}
		\caption{}\label{subfig:UVmodelssecondHiggsandscalar}
	\end{subfigure}\\
	% \vspace*{0.25cm}
	\begin{subfigure}[b]{0.45\linewidth}\centering
		\begin{tikzpicture}
			\begin{feynman}
				\vertex (lamDelta1) at (0,0);
				\vertex (phi1) at (-1,-1.8) {\(\phi\)};
				\vertex (lamEDelta1) at (1.5,0);
				\vertex (phi2) at (1.5,-1.8) {\(\phi^*\)};
				\vertex (lamE) at (3,0);
				\vertex (phi3) at (4,-1.8) {\(\phi\)};
				\vertex (eR) at (-1,1.8) {\(e_R\)};
				\vertex (lL) at (4,1.8) {\(\bar{l}_L\)};
				\diagram*{
					(phi1) -- [scalar] (lamDelta1);
					(phi2) -- [scalar] (lamEDelta1);
					(phi3) -- [scalar] (lamE);
					(eR) -- [fermion] (lamDelta1) -- [fermion, red, edge label = \(\Psi_1\)] (lamEDelta1) -- [fermion, red, edge label = \(\Psi_2\)] (lamE) -- [fermion] (lL);
				};
			\end{feynman}			
		\end{tikzpicture}
		\caption{}\label{subfig:UVmodelsVLLpairs}
	\end{subfigure}
	\begin{subfigure}[b]{0.45\linewidth}\centering
		\begin{tikzpicture}
			\begin{feynman}
				\vertex (uR) at (0,0) {\(e_R\)};
				\vertex[below right= of uR] (v1);
				\vertex[right= of v1] (v2);
				\vertex[above right= of v2] (phi) {\(\phi\)};
				\vertex[below right= of v2] (phistar) {\(\phi^*\)};
				\vertex[below= of v1] (v3);
				\vertex[below left= of v3] (qL) {\(\bar{\ell}_L\)};
				\vertex[below right = of v3] (phistar2) {\(\phi\)};
				\diagram*{
					(uR) -- [fermion] (v1) -- [fermion, red, edge label = \(\Psi\)] (v3) -- [fermion] (qL);
					(v1) -- [scalar, red, edge label = \(\Phi\)] (v2) -- [scalar] (phi);
					(v2) -- [scalar] (phistar);
					(v3) -- [scalar] (phistar2);
				};
			\end{feynman}
		\end{tikzpicture} 
		\caption{}\label{subfig:UVmodelsVLLandscalar}
	\end{subfigure}
	\caption{Examples of Feynman diagrams showing how new states can generate an operator of type $\bar{l}_{L} \phi e_{R} \,(\phi^{\dagger}\phi)$ at tree level. \textbf{(a)} A scalar doublet $\varphi$. \textbf{(b)} A pair of scalars $\varphi$ and $\Phi$. \textbf{(c)} A pair of vector-like leptons $\Psi_1$ and $\Psi_2$. \textbf{(d)} A new scalar $\Phi$ and a vector-like lepton $\Psi$.}
	\label{fig:UVmodel}
\end{figure}

The first deviation from this picture is observed if two new mediators are simultaneously present. The reason is that the NP effect does not have to proceed through the insertion of the renormalisable Yukawa coupling as shown in Fig.~\ref{fig:UVmodel} and the chiral enhancement of $v/m_e$ can be achieved. Therefore, we have:

\begin{itemize}
	\item Models involving only scalar fields. In particular, a model can involve either only an $SU(2)$ doublet $\varphi$, or $\varphi$ in combination with an additional scalar field (either an $SU(2)$ singlet or a triplet). The mechanism by which new scalars could give rise to $\cO_{e\phi}$ is shown in Figs.~\ref{subfig:UVmodelssecondHiggs} and~\ref{subfig:UVmodelssecondHiggsandscalar}; 
	\item Models with two additional representations of VLLs, as shown in Fig.~\ref{subfig:UVmodelsVLLpairs};
	\item Models with a  VLL and a scalar. The associated Feynman diagram is shown in  Fig.~\ref{subfig:UVmodelsVLLandscalar}.
\end{itemize}
In all cases, the complete NP Lagrangian takes the form
\begin{equation}
    \cL_{\rm NP} = \cL_{\rm SM} + \cL_{\rm quad} + \cL_{\rm int}\,.
    \label{eq:NPLagrangiangeneral}
\end{equation}
The second term involves the kinetic and massive terms for the NP particles. In the presence of a new scalar (generically denoted by $\Phi$), the quadratic term is
\begin{equation}
    \cL_{\rm quad}^\Phi = \eta_\Phi\lsquare\left(D_\mu\Phi\right)^\dagger \, \left(D^\mu \Phi\right) - M^2_\Phi \Phi^\dagger \, \Phi\,\rsquare\, , \quad \eta_\Phi = \begin{cases}
        1 \quad \text{complex representation}\\
        \frac{1}{2} \quad \text{real representation}
    \end{cases}\,.
\end{equation}
If the NP particle is a VLL $\Psi$, then the quadratic term reads
\begin{equation}
    \cL_{\rm quad}^\Psi = \eta_\Psi \lsquare\bar{\Psi} i \gamma_\mu D^\mu \Psi - M_\Psi\bar{\Psi}\Psi\,\rsquare\,, \quad 
    \eta_\Psi = \begin{cases}
       1 \quad \text{Dirac field}\\
       \frac{1}{2} \quad \text{Majorana field}
    \end{cases}.
\end{equation}
As for the third term in Eq.~\eqref{eq:NPLagrangiangeneral}, it contains the interaction terms of the NP particles. Though the explicit interaction Lagrangians for each type of model will be detailed in the following, we point out that for all models we assume only one generation of NP particles coupled to leptons.

We indicate the representation of the new states under the SM gauge group as follows: $(SU(3), SU(2))_Y$ with $Y=Q-T_3$ and $Q$ the electric charge and $T_3$ the quantum number under the third $SU(2)$ generator $T_3=\sigma_3/2$. States with $Y=0$ are understood to be real.

% % % % % % % % % % % % 
\subsection{Models with scalars}
\label{subsec:models_scalars}
% % % % % % % % % % % %

Only one single-particle extension of the SM results in the chirally enhanced $\kappa_e$. It is a scalar doublet $\varphi\sim(\mathbf{1},\mathbf{2})_{1/2}$ that gives rise to $\cO_{e\phi}$ through the diagram shown in Fig.~\ref{subfig:UVmodelssecondHiggs}. The relevant interaction Lagrangian reads
\begin{equation}
    -\cL_\varphi = \lsquare y^e_{\varphi}\rsquare_{ij} \varphi^\dagger\, \bar{e}_{R}^i\,l_{L}^j + \lambda_\varphi\left(\varphi^\dagger\phi\right)\left(\phi^\dagger\phi\right) + \hc\,, \label{eq:L2HDM}
\end{equation}
and induces $\cO_{e\phi}$ after $\varphi$ is integrated out. The associated Wilson coefficient is 
\begin{equation}
    \lsquare\cC_{e\phi}\rsquare_{ij} = \frac{\lambda_\phi \lsquare y^e_\varphi\rsquare^*_{ji}}{M^2_\varphi}\,  \Longrightarrow \kappa_e = 1 - \frac{v^3}{\sqrt{2}m_e M_\varphi^2}\lambda_\varphi \lsquare y^e_\varphi\rsquare^*_{11}\,. \label{eq:cephi_2hdm}
\end{equation}
In Eq.~\eqref{eq:L2HDM}, we have omitted further interactions among the scalar fields as they are irrelevant to our discussion.
Furthermore, there are three additional models with the combination of $\varphi$ and an extra scalar which can be either a singlet or a triplet of $SU(2)$. We collect the new fields with their transformation properties under $G_{\rm SM}$ in Tab.~\ref{tab:newphysicsscalars}. The interaction Lagrangians for each of the scalars with the SM particles are
\begin{equation}
    -\cL_{S} = \kappa_S S\phi^\dagger\phi + \lambda_S S S \phi^\dagger \phi + \kappa_{S^3} S S S\,, \label{eq:LagS}
\end{equation}
\begin{equation}
    -\cL_\Xi = \kappa_\Xi\, \phi^\dagger \Xi^I \sigma^I \phi	+ \lambda_\Xi \left(\Xi^I \Xi^I\right) \left(\phi^\dagger \phi\right)\,, \label{eq:Lagtrip1}
\end{equation}
\vspace{-0.6cm}
\begin{align}
    -\cL_{\Xi_1} &=  \frac{1}{2} \lambda_{\Xi_1} \left(\Xi^{I\dagger}_1 \Xi^I_1\right) \, \left(\phi^\dagger\phi\right)  + \frac{1}{2} \lambda'_{\Xi_1} f_{IJK}  \left(\Xi_1^{I\dagger} \Xi_1^J\right)	\,\left(\phi^\dagger \sigma^K \phi\right)	+\nonumber\\
                 &\quad+\left\{ \lsquare y_{\Xi_1}\rsquare_{rs} \Xi^{I\dagger}_1 \bar{l}_{Lr} \sigma^I i\sigma_2 l^c_{Ls}
		                 	+ \kappa_{\Xi_1} \Xi_1^{I\dagger} \left(\tilde{\phi}^\dagger \sigma^I \phi\right)
			                + \hc\right\}\,. \label{eq:LagTrip2}
\end{align}
With these definitions, the complete Lagrangian densities are given by the sum of the individual contributions and the mixed terms, as detailed below:
\begin{itemize}
	\item Model S1: $\varphi + S$
		\begin{equation}
			-\cL_{\rm S1} = - \cL_\varphi -\cL_{S}  + \left( \kappa_{S\varphi} S \varphi^\dagger \phi + \hc\right)\,,
		\end{equation}
	\item Model S2: $\varphi + \Xi$
		\begin{equation}
			-\cL_{\rm S2} = - \cL_\varphi - \cL_\Xi + \left(\kappa_{\Xi\varphi}\, \Xi^I \left(\varphi^\dagger \sigma^I \phi\right)+ \hc\right)\,,
		\end{equation}
	\item Model S3: $\varphi + \Xi_1$
		\begin{equation}
			-\cL_{\rm S3} = - \cL_\varphi - \cL_{\Xi_1} + \left(\kappa_{\Xi_1\varphi} \, \Xi^{I\dagger}_1 \left(\tilde{\varphi}^\dagger \sigma^I  \phi\right) + \hc\right)\,.
		\end{equation}
\end{itemize}
We note that in the equations above we did not consider interaction terms in the scalar potential where in Eqs.~\eqref{eq:LagS}--\eqref{eq:LagTrip2} the SM-like doublet field $\phi$ is replaced by $\varphi$. Those do not impact our study, so we omit them. 

The tree-level matching to $\cO_{e\phi}$ is presented in Tab.~\ref{tab:Cephi_scalarpairs}. We point out that these results can be split into the sum of two terms, one solely due to the scalar doublet $\varphi$ and another involving both NP scalars.

\begin{table}[t]
	\centering
	{\renewcommand{\arraystretch}{1.8}
		\begin{tabular}{|c|cccc|}
			\hline
			Scalars	& 	$S$		&	$\varphi$	&	$\Xi$	&	$\Xi_1$ \\
			\hline
			Irrep.	  &  $\left(\bf{1}, \bf{1}\right)_0$	& $\left(\bf{1},\bf{2}\right)_\frac{1}{2}$	&$\left(\bf{1},\bf{3}\right)_0$	&	$\left(\bf{1},\bf{3}\right)_1$		\\
                         % &   Real   &     Complex   &     Real      & Complex \\
			\hline
		\end{tabular}
	}
	\caption{The NP scalars studied in the context of enhanced electron Yukawa coupling.
    }
	\label{tab:newphysicsscalars}
\end{table}
% % % % % % % % % % % %

\begin{table}[t]
    \centering
    {\renewcommand{\arraystretch}{1.7}
    \begin{tabular}{|c|c|}
    \hline
    Model     &   $\lsquare \cC_{e\phi}\rsquare_{11}$\\
    \hline
    S1     & $\frac{1 }{M^2_\varphi} \lround\lambda_\varphi-\frac{1}{M_S^2}\, \kappa_{S\varphi}\kappa_S \rround \lsquare y^e_\varphi\rsquare^*_{11}$ \\
    S2     & $\frac{1 }{M^2_\varphi} \lround\lambda_\varphi-\frac{1}{ M^2_\Xi} \,\kappa_{\Xi\varphi} \kappa_\Xi \rround \lsquare y^e_\varphi \rsquare^*_{11}$ \\
    S3     & $\frac{1 }{M^2_\varphi}\lround \lambda_\varphi-\frac{2}{M_{\Xi_1}^2} \, \kappa^*_{\Xi_1\varphi} \kappa_{\Xi_1} \rround \lsquare y^e_\varphi\rsquare^*_{11}$\\
    \hline
    \end{tabular}
    }
    \caption{Tree-level matching for models with pairs of scalars to the $\cO_{e\phi}$ SMEFT operator.}
    \label{tab:Cephi_scalarpairs}
\end{table}

%%%%%%%%%%%%%%%%%%%%%%%%%%%%%%%%
\subsection{Models with vector-like lepton pairs}
\label{subsec:models_vllpairs}
%%%%%%%%%%%%%%%%%%%%%%%%%%%%%%%%

There are five simplified models containing pairs of the vector-like leptons listed in Tab.~\ref{tab:vectorlikeleptons} that generate a contribution to the $\cO_{e\phi}$ operator, unsuppressed by the renormalisable Yukawa coupling $y_e$. Schematically the interaction Lagrangian reads
\begin{equation}
    -\cL_{\rm 2VLL} \supset \lambda_L \bar{L}_1\phi l_L + \lambda_R \bar{L}_2\phi l_R + \lambda_{12}\bar{L}_1\phi L_2 \, \Rightarrow \, \kappa_e = 1 + \alpha \frac{v^3\lambda_L \lambda_R^* \lambda_{12}}{m_e M_1 M_2}\,,  
\end{equation}
where $\alpha$ is a model-dependent number resulting from the tree-level matching. In the following, we list all the possible models involving pairs of VLLs:
\begin{itemize}
	\item Model L1: $E+\Delta_1$
	\begin{equation}
		-\cL_{\rm L1} = \lsquare\lambda_E\rsquare_r \bar{E}_R \phi^\dagger l_{Lr} + \lsquare\lambda_{\Delta_1}\rsquare_r \bar{\Delta}_{1L} \phi e_{Rr} + \lambda_{E\Delta_1} \bar{E}_L \phi^\dagger \Delta_{1R} +\hc \,,
        \label{eq:L1int}
	\end{equation}
	\item Model L2: $E+\Delta_3$
	\begin{equation}
		-\cL_{\rm L2} = \lsquare\lambda_E\rsquare_r \bar{E}_R \phi^\dagger l_{Lr} + \lsquare\lambda_{\Delta_3}\rsquare_r \bar{\Delta}_{3L} \tilde{\phi} e_{Rr}  + \lambda_{E \Delta_3} \bar{E}_{Lr} \tilde{\phi}^\dagger \Delta_{3R} + \hc\,,
	\end{equation}
        \item Model L3: $\Sigma_1 + \Delta_3$
	\begin{equation}
		-\cL_{\rm L3} =  \frac{1}{2} \lsquare\lambda_{\Sigma_1}\rsquare_r \bar{\Sigma}^I_{1R} \phi^\dagger \sigma^I l_{Lr} +  \lsquare\lambda_{\Delta_3}\rsquare_r \bar{\Delta}_{3L} \tilde{\phi} e_{Rr} +  \frac{1}{2} \lambda_{\Sigma_1 \Delta_3}	\bar{\Sigma}^I_{1L} \tilde{\phi}^\dagger \sigma^I \Delta_{3R} + \hc\,,
	\end{equation}
	\item Model L4: $\Sigma_1 + \Delta_1$ 
	\begin{equation}
		-\cL_{\rm L4} =  \frac{1}{2} \lsquare\lambda_{\Sigma_1}\rsquare_r \bar{\Sigma}^I_{1R} \phi^\dagger \sigma^I l_{Lr} + \lsquare\lambda_{\Delta_1}\rsquare_r \bar{\Delta}_{1L} \phi e_{Rr}  + \frac{1}{2} \lambda_{\Sigma_1 \Delta_1} \bar{\Sigma}^I_{1L} \phi^\dagger \sigma^I \Delta_{1R} +\hc \,,
	\end{equation}
    \item Model L5: $\Sigma + \Delta_1$ 
	\begin{equation}
		-\cL_{\rm L5} =  \frac{1}{2} \lsquare\lambda_{\Sigma}\rsquare_r \bar{\Sigma}^I_R \tilde{\phi}^\dagger \sigma^I l_{Lr} + \lsquare\lambda_{\Delta_1}\rsquare_r \bar{\Delta}_{1L} \phi e_{Rr}  + \frac{1}{2} \lambda_{\Sigma \Delta_1} \bar{\Sigma}^{ I\, c }_{R} \tilde{\phi}^\dagger \sigma^I \Delta_{1R} + \hc \,.
        \label{eq:L5int}
	\end{equation}
\end{itemize}

\begin{table}[t]
	\centering
	{\renewcommand{\arraystretch}{1.8}
	\begin{tabular}{|c|ccccc|}
		\hline
		VLL	& $E$	& $\Delta_1$	&	$\Delta_3$	&	$\Sigma$ 	&	$\Sigma_1$ \\
		\hline
		Irrep.	 & $\left(\bf{1},\bf{1}\right)_{-1}$ & $\left(\bf{1},\bf{2}\right)_{-\frac{1}{2}}$ & $\left(\bf{1},\bf{2}\right)_{-\frac{3}{2}}$ & $\left(\bf{1},\bf{3}\right)_{0}$ & $\left(\bf{1},\bf{3}\right)_{-1}$\\
                    % & D & D & D & M & D \\
		\hline
	\end{tabular}
	}
	\caption{The vector-like leptons studied in the context of enhanced electron Yukawa coupling.}
	\label{tab:vectorlikeleptons}
\end{table}
% % % % % % % % % % % %
The expression of $\cC_{e\phi}$ found when studying these models can be split into two contributions: the first involves only one VLL at a time, while the second is due to the presence of both VLLs (as shown in Fig.~\ref{subfig:UVmodelsVLLpairs}). Considering the specific example of Model L1, the expression reads
\begin{equation}
    \lsquare\cC_{e\phi}\rsquare_{11} = \left(\frac{\lsquare y_e^*\rsquare_{11} \, |\lsquare\lambda_E\rsquare_1|^2}{2M_E^2} + \frac{\lsquare y_e^*\rsquare_{11}\, |\lsquare\lambda_{\Delta_1}\rsquare_1|^2}{2M_{\Delta_1}^2}\right) + \left(-\frac{\lambda_{E\Delta_1} \lsquare\lambda_E^*\rsquare_1 \lsquare\lambda_{\Delta_1}\rsquare_1}{M_E M_{\Delta_1}}\right)\,.
\end{equation}
As discussed above, the $y_e$ pieces are subleading due to the requirement to tune the electron mass and pass the electroweak precision tests. Accordingly, in Tab.~\ref{tab:Cephi_vllpairs}, we provide the expressions for the dominant contribution stemming from the mixing of the two heavy states.

\begin{table}[t]
    \centering
    {\renewcommand{\arraystretch}{1.7}
    \begin{tabular}{|c|c||c|c|}
    \hline
    Model & $\lsquare \cC_{e\phi} \rsquare_{11}$ & Model & $\lsquare \cC_{e\phi} \rsquare_{11}$ \\
    \hline
    L1    & $-\frac{1}{M_E M_{\Delta_1}} \,\lambda_{E\Delta_1} \lsquare \lambda_E\rsquare_1^* \lsquare\lambda_{\Delta_1}\rsquare_1$                 & L4    & $-\frac{1}{4M_{\Sigma_1} M_{\Delta_1}} \,\lambda_{\Sigma_1\Delta_1} \lsquare \lambda_{\Sigma_1}\rsquare_1^* \lsquare\lambda_{\Delta_1}\rsquare_1$ \\
    L2    & $-\frac{1}{M_E M_{\Delta_3}}\, \lambda_{E\Delta_3} \lsquare \lambda_E\rsquare_1^* \lsquare\lambda_{\Delta_3}\rsquare_1$                 & L5    & $-\frac{1}{2M_\Sigma M_{\Delta_1}} \,\lambda_{\Sigma\Delta_1} \lsquare \lambda_\Sigma\rsquare_1^* \lsquare\lambda_{\Delta_1}\rsquare_1$\\
    L3    & $\frac{1}{4M_{\Sigma_1} M_{\Delta_3}}\, \lambda_{\Sigma_1\Delta_3} \lsquare \lambda_{\Sigma_1}\rsquare_1^* \lsquare\lambda_{\Delta_3}\rsquare_1$ &     &  \\
    \hline
    \end{tabular}
    }
    \caption{Dominant contribution to the electron Yukawa for models with pairs of VLLs. }
    \label{tab:Cephi_vllpairs}
\end{table}

%%%%%%%%%%%%%%%%%%
\subsection{Models with a vector-like lepton and a scalar}
\label{subsec:models_vllplusscalar}
%%%%%%%%%%%%%%%%%%

The final class of models that will be considered involves a combination of VLLs from Tab.~\ref{tab:vectorlikeleptons} and scalars from Tab.~\ref{tab:newphysicsscalars}. The models with mixed interactions between a VLL and a new scalar are six in total and are labeled as follows:
\begin{itemize}
	\item Model SL1: $S+E$
	\begin{equation}
		-\cL_{\rm SL1}= -\cL_S + \lround \lsquare\lambda_E\rsquare_r \bar{E}_R \phi^\dagger l_{Lr} + \lsquare\lambda_{SE}\rsquare_r S \bar{E}_L e_{Rr}  +\hc \rround \,,
	\end{equation}
	\item Model SL2: $S+\Delta_1$
	\begin{equation}
		-\cL_{\rm SL2}= -\cL_S +\lround \lsquare\lambda_{\Delta_1}\rsquare_r \bar{\Delta}_{1L} \phi e_{Rr} + \lsquare \lambda_{\mathcal{S}\Delta_1} \rsquare_r S \,\bar{\Delta}_{1R}\, l_{Lr}  + \hc \rround \,,
	\end{equation}
	\item Model SL3: $\Xi + \Delta_1$
	\begin{equation}
		-\cL_{\rm SL3} = -\cL_\Xi + \lround\lsquare\lambda_{\Delta_1}\rsquare_r \bar{\Delta}_{1L} \phi e_{Rr} + \lsquare\lambda_{\Xi \Delta_1}\rsquare_r \Xi^I \bar{\Delta}_{1R} \sigma^I l_{Lr} + \hc\rround \,,
	\end{equation}
	\item Model SL4: $\Xi + \Sigma_1$
	\begin{equation}
		-\cL_{\rm SL4} = -\cL_\Xi +  \lround\frac{1}{2} \lsquare\lambda_{\Sigma_1}\rsquare_r \bar{\Sigma}^I_{1R} \phi^\dagger \sigma^I l_{Lr} + \lsquare\lambda_{\Xi \Sigma_1}\rsquare_r \Xi^I \bar{\Sigma}_{1L}^I e_{Rr} + \hc \rround\,,
	\end{equation}
	\item Model SL5: $\Xi_1 + \Delta_3$
	\begin{equation}
		-\cL_{\rm SL5} = -\cL_{\Xi_1} + \lround\lsquare\lambda_{\Delta_3}\rsquare_r \bar{\Delta}_{3L} \tilde{\phi} e_{Rr} + \lsquare\lambda_{\Xi_{1}\Delta_3}\rsquare_r \Xi_1^{I\dagger} \bar{\Delta}_{3R} \sigma^I l_{Lr} +\hc\rround\,,
	\end{equation}
	\item Model SL6: $\Xi_1 + \Sigma$
	\begin{equation}
		-\cL_{\rm SL6} = -\cL_{\Xi_1} + \lround\frac{1}{2} \lsquare\lambda_{\Sigma}\rsquare_r \bar{\Sigma}^I_R \tilde{\phi}^\dagger \sigma^I l_{Lr}  + \lsquare\lambda_{\Xi_{1}\Sigma}\rsquare_r \Xi_1^{I\dagger} \bar{\Sigma}_R^{c\, I} e_{Rr}^{c} + \hc\rround\,.
	\end{equation} 
\end{itemize}

Before discussing further, we point out that additional interactions involving only NP particles are possible. In particular, one can write Lagrangian terms involving two VLLs and one new scalar. However, we chose to neglect them as they do not affect the phenomenology we are interested in.

The results for the tree-level matching for each of these models to the operator $\cO_{e \phi}$ are presented in Tab.~\ref{tab:Cephi_scalar+VLL}. The reported terms are the dominant ones, independent of the renormalisable Yukawa coupling $y_e$.

\begin{table}[t]
\centering
    {\renewcommand{\arraystretch}{1.7}
    \begin{tabular}{|c|c||c|c|}
    \hline
    Model   &   $\lsquare \cC_{e\phi}\rsquare_{11}$  &   Model   &   $\lsquare \cC_{e\phi}\rsquare_{11}$ \\
    \hline
    SL1     &  $-\frac{1}{M_E M_S^2} \,\lsquare\lambda_{SE}\rsquare_1 \kappa_S \lsquare\lambda_E \rsquare_1^*$ & SL4 & $-\frac{1}{2M_\Xi^2 M_{\Sigma_1}} \,\lsquare \lambdaNPNP{\Xi}{\Sigma_1}\rsquare_1 \kappa_\Xi \lsquare \lambda_{\Sigma_1}\rsquare^*_1$ \\
    SL2     &  $-\frac{1}{M_S^2 M_{\Delta_1}} \,\lsquare \lambdaNPNP{S}{\Delta_1}\rsquare_1^* \kappa_S \lsquare \lambda_{\Delta_1}\rsquare_1$     &   SL5 & $-\frac{2}{M_{\Xi_1}^2 M_{\Delta_3}} \, \lsquare \lambdaNPNP{\Xi_1}{\Delta_3}\rsquare_1^* \kappa_{\Xi_1} \lsquare \lambda_{\Delta_3}\rsquare_1$ \\
    SL3     &  $-\frac{1}{M_\Xi^2 M_{\Delta_1}} \,\lsquare \lambdaNPNP{\Xi}{\Delta_1}\rsquare_1^* \kappa_\Xi \lsquare \lambda_{\Delta_1}\rsquare_1$     &   SL6 & $-\frac{1}{M_{\Xi_1}^2 M_{\Sigma}} \,\lsquare \lambdaNPNP{\Xi_1}{\Sigma}\rsquare_1^* \kappa_{\Xi_1} \lsquare \lambda_{\Sigma}\rsquare^*_1$ \\
    \hline
    \end{tabular}
    }
    \caption{Dominant contribution to the modification of the electron Yukawa for models involving one VLL and one scalar.}
    \label{tab:Cephi_scalar+VLL}
\end{table}

%%%%%%%%%%%%%%%%%%%%%%%%%%%%%%%%%%
\subsection{Generated SMEFT operators}
%%%%%%%%%%%%%%%%%%%%%%%%%%%%%%%%%%

In addition to the $\cO_{e\phi}$ operator, each of the above models generates other operators once the heavy NP particles have been integrated out. This is especially relevant as those will be constrained from flavour, Higgs, and electroweak precision data. In this work, we consider operators up to dimension six generated at the tree and one-loop level. 
Further, the models containing the scalar $\Xi_1~\sim({\bf{1}},{\bf{3}})_{1}$ or the vector-like lepton $\Sigma\sim({\bf{1}},{\bf{3}})_0$  generate the dimension-five operator at the tree-level
\begin{equation}
   \lsquare\cO_{\nu\nu}\rsquare_{ij} = \left(\tilde{\phi}^\dagger l_L^i\right)^T\,C\left(\tilde{\phi}^\dagger l_L^j\right)\,, \quad C = i \gamma^2\gamma^0\,.
\end{equation}
Besides the expressions for the tree-level matching to the operator $\cO_{e\phi}$, which was presented for each model in Eq.~\eqref{eq:cephi_2hdm} and Tabs.~\ref{tab:Cephi_scalarpairs}--\ref{tab:Cephi_scalar+VLL}, we also report in Tabs.~\ref{tab:treelevelmatching_novarphi}--\ref{tab:treelevelmatching_varphi} the tree-level matching results to the other operators that will enter various constraints. At the tree level, each particle is associated with a specific set of operators, allowing to organise the results depending on the NP particle and not the specific model, i.e.~$\cC_{e\phi}$ is the only coefficient that depends on the coupling between two NP particles. Besides the operators already defined, the following additional operators appear in Tabs.~\ref{tab:treelevelmatching_novarphi}--\ref{tab:treelevelmatching_varphi}
\begin{align}
    \cO_\phi &= (\phi^\dagger \phi)^3\,,
    \label{eq:Ophi}\\
    \lsquare\cO_{ll}\rsquare
    _{ijkl}&= \lround\bar{l}_L^i\gamma_\mu l_L^j\rround\lround\bar{l}_L^k\gamma^\mu l_L^l\rround \,,\\
    \lsquare\cO_{le}\rsquare_{ijkl} &= \lround\bar{l}_L^i\gamma_\mu l_L^j\rround\lround\bar{e}_R^k\gamma^\mu e_R^l\rround\,.
\end{align}
As for the one-loop matching coefficients, we have used both \texttt{Matchete}~\cite{Fuentes-Martin:2022jrf} and \texttt{SOLD}~\cite{Guedes:2023azv,Guedes:2024vuf}, and checked that their results agree. We provide the expressions for the one-loop matching in the equal mass limit in the ancillary notebook attached to this work.

% % % % % % % % % % % % 
\section{Constraints}
\label{sec:constraints}
% % % % % % % % % % % % 

The signatures of the underlying dynamics responsible for modifying the electron Yukawa coupling could be seen in different processes involving the first-generation leptons, electroweak gauge, and the Higgs bosons. In the following, we study various constraints on the simplified models outlined in the previous section. 

\begin{landscape}
    \begin{table}[t]
	\centering
    {\renewcommand{\arraystretch}{1.7}
	\begin{tabular}{|c|c|c|ccc|ccc|c|}
        \hline
	 	       & $d=4$	&	$d=5$&	\multicolumn{7}{c|}{$d=6$}\\
        \cline{2-10}
	           & $\cC_{\phi^4}$ & $\lsquare\cC_{\nu\nu}\rsquare_{ij}$ & $\cC_{\phi}$ & $\cC_{\phi D}$ & $\cC_{\phi\Box}$ & $\lsquare\cC_{\phi e}\rsquare_{ij}$ & $\lsquare\cC^{(1)}_{\phi l}\rsquare_{ij}$ & $\lsquare\cC^{(3)}_{\phi l}\rsquare_{ij}$ & $\lsquare\cC_{ll}\rsquare_{ijkl}$ \\
        \hline\hline
        $S$       & $ \frac{|\kappa_S|^2}{2M^2_S}$  & - & $-\frac{\kappa_S^2}{M_S^4}\lround \lambda_S - \frac{\kappa_S\kappa_{S_3}}{M^2_S}\rround$ & - & $\-\frac{\kappa_S^2}{2M_S^4}$ & - & - & - & - \\
        $\Xi$     & $ \frac{\kappa_\Xi^2}{M_\Xi^2}\lround\frac{1}{2} - \frac{2 \mu^2 }{M_\Xi^2}\rround$ & - & $\frac{\kappa_\Xi^2}{M^4_\Xi}\lround4\lambda -\lambda_\Xi\rround$ & $-\frac{2\kappa_\Xi^2}{M_\Xi^4}$ & $\frac{\kappa_\Xi^2}{2M_\Xi^4}$ & - & - & - & - \\
        $\Xi_1$   & $ \frac{|\kappa_{\Xi_1}|^2}{M_{\Xi_1}^2}\lround\frac{1}{2} - \frac{4\mu^2}{M_\Xi^2}\rround$ & $-\frac{2\kappa_{\Xi_1} \lsquare y_{\Xi_1}\rsquare_{ji}}{M^2_{\Xi_1}}$ & $\frac{|\kappa_{\Xi_1}|^2}{M^4_{\Xi_1}}\lround 8\lambda-2\lambda_{\Xi_1} +\sqrt{2}\lambda_{\Xi_1}' \rround$ & $\frac{4|\kappa_{\Xi_1}|^2}{M_{\Xi_1}^4}$ & $\frac{2|\kappa_{\Xi_1}|^2}{M_{\Xi_1}^4}$ & - & - & - & $\frac{\lsquare y_{\Xi_1}\rsquare_{ki} \lsquare y^*_{\Xi_1} \rsquare_{lj}}{M^2_{\Xi_1}}$ \\
        \hline\hline
        $E$          & - & - & - & - & - & - & $ - \frac{\lsquare\lambda_E\rsquare_j\lsquare\lambda_E^*\rsquare_i}{4M_E^2}$ & $ - \frac{\lsquare\lambda_E\rsquare_j\lsquare\lambda_E^*\rsquare_i}{4M_E^2}$ & - \\
        $\Delta_1$   & - & - & - & - & - & $\frac{\lsquare \lambda_{\Delta_1} \rsquare_j \lsquare \lambda_{\Delta_1}^* \rsquare_i}{2 M_{\Delta_1}^2}$   & - & - & - \\
        $\Delta_3$   & - & - & - & - & - & $-\frac{\lsquare \lambda_{\Delta_3} \rsquare_j \lsquare \lambda_{\Delta_3} \rsquare^*_i}{2 M_{\Delta_3}^2}$  & - & - & - \\
        $\Sigma$     & - & $\frac{\lsquare \lambda_{\Sigma} \rsquare_i \lsquare \lambda_{\Sigma} \rsquare_j}{8M_\Sigma}$  & - & - & - & - & $ \frac{3\lsquare\lambda_{\Sigma}\rsquare_j\lsquare\lambda_{\Sigma}^*\rsquare_i}{16M_{\Sigma}^2}$ & $ \frac{\lsquare\lambda_{\Sigma}\rsquare_j\lsquare\lambda_{\Sigma}^*\rsquare_i}{16M_{\Sigma}^2}$ & - \\
        $\Sigma_1$   & - & - & - & - & - & - &  $ -\frac{3\lsquare\lambda_{\Sigma_1}\rsquare_j\lsquare\lambda_{\Sigma_1}\rsquare^*_i}{16M_{\Sigma_1}^2}$ &  $ \frac{\lsquare\lambda_{\Sigma_1}\rsquare_j\lsquare\lambda_{\Sigma_1}\rsquare^*_i}{16M_{\Sigma_1}^2}$ & - \\
        \hline    
		\end{tabular}
    } 
    \caption{Tree-level generated operators by each of the NP particles introduced in Tabs.~\ref{tab:newphysicsscalars}--\ref{tab:vectorlikeleptons}, except for $\varphi$. The operators are separated based on their dimensions. The dimension four operator renonormalises the SM four Higgs interaction term \cite{deBlas:2017xtg}.}
    \label{tab:treelevelmatching_novarphi}
\end{table}

\begin{table}[b]
    \centering
    {\renewcommand{\arraystretch}{1.7}
    \begin{tabular}{|ccccccccc|}
    \hline
    $\lsquare\cC_{le}\rsquare_{ijkl}$ & $\lsquare\cC_{qu}^{(1)}\rsquare_{ijkl}$ & $\lsquare\cC_{qu}^{(8)}\rsquare_{ijkl}$ & $\lsquare\cC_{qd}^{(1)}\rsquare_{ijkl}$ & $\lsquare\cC^{(8)}_{qd}\rsquare_{ijkl}$ & $\lsquare\cC_{ledq}\rsquare_{ijkl}$ & $\lsquare\cC_{quqd}^{(1)}\rsquare_{ijkl}$ & $\lsquare\cC^{(1)}_{lequ}\rsquare_{ijkl}$ & $\cC_\phi$ \\
    \hline\hline
    $-\frac{\lsquare y^e_\varphi\rsquare^*_{li} \lsquare y^e_\varphi\rsquare_{kj}}{2M^2_\varphi}$ & $-\frac{\lsquare y^u_\varphi\rsquare^*_{jk} \lsquare y^u_\varphi\rsquare_{il}}{6M^2_\varphi}$ & $-\frac{\lsquare y^u_\varphi\rsquare^*_{jk} \lsquare y^u_\varphi\rsquare_{il}}{M^2_\varphi}$ & $-\frac{\lsquare y^d_\varphi\rsquare^*_{li} \lsquare y^d_\varphi\rsquare_{kj}}{6M^2_\varphi}$ & $-\frac{\lsquare y^d_\varphi\rsquare^*_{li} \lsquare y^d_\varphi\rsquare_{kj}}{M^2_\varphi}$ & $\frac{\lsquare y^d_\varphi \rsquare_{kl} \lsquare y^e_\varphi \rsquare_{ji}^*}{M^2_\varphi}$ & $\frac{\lsquare y^u_\varphi \rsquare_{ij} \lsquare y^d_\varphi \rsquare_{lk}^*}{M^2_\varphi}$ & $\frac{\lsquare y^u_\varphi \rsquare_{kl} \lsquare y^e_\varphi \rsquare_{ji}^*}{M^2_\varphi}$ & $\frac{|\lambda_\varphi|^2}{M^2_\varphi}$\\
    \hline
    \end{tabular}
    }
    \caption{Tree-level matching for the operators generated by integrating out the heavy $\varphi$ NP particle. Under our assumption that $\varphi$ only couples to leptons (i.e., $y_\varphi^u=y_\varphi^d=0$), only the rightmost and leftmost SMEFT coefficients remain.}
    \label{tab:treelevelmatching_varphi}
\end{table}
\end{landscape}

\subsection{Flavour Physics}
\label{subsec:flavourphysics}

\subsubsection{Models with vector-like leptons}

New vector-like leptons exhibit a potentially rich flavour structure through the Yukawa couplings with the SM leptons and the Higgs. In a generic setting, these new particles break all accidental lepton flavour symmetries leading to processes that violate lepton number and flavour. 

\paragraph{Lepton number violation.} Lepton number is an accidental symmetry of the SM violated already at dimension $d=5$ in the SMEFT expansion~\cite{Weinberg:1979sa}. Following electroweak symmetry breaking, the operator
\begin{equation}
    \mathcal{L}_{\rm SMEFT} \supset [\mathcal{C}_{\nu\nu}]_{ij}\, \bar{\ell}_L^{c,i} \tilde{\phi}^*\tilde{\phi}^\dagger \ell_L^j\,,
    \label{eq:Weinberg_op}
\end{equation}
generates Majorana neutrino masses and contributes to neutrinoless double-beta decay ($0\nu\beta\beta$). The absence of experimental evidence for the transition $(A,Z)\to (A,Z+2)$  characterized by two same-sign electrons and no neutrinos, imposes strict limits on $[\mathcal{C}_{\nu\nu}]_{11}$, specifically $[\mathcal{C}_{\nu\nu}]_{11}\lesssim 10^{-14}$ GeV$^{-1}$~\cite{GERDA:2020xhi}. This severely constrains models predicting nonzero $[\mathcal{C}_{\nu\nu}]_{11}$, necessitating new physics at mass scales so high that any impact on the electron Yukawa coupling is negligible.

In models capable of enhancing $\kappa_e$, two states generate the operator in Eq.~\eqref{eq:Weinberg_op}; a scalar $\Xi_1 \sim (\mathbf{1},\mathbf{3})_1$, and a vector-like fermion $\Sigma \sim (\mathbf{1},\mathbf{3})_0$ associated with the Type II and Type III seesaw mechanisms, respectively~\cite{Zee:1980ai,Foot:1988aq}. While fine-tuning could, in principle, allow $M_{\Xi_1}$ and $M_{\Sigma}$ to lie at the TeV scale, this is not technically natural for $\Xi_1$ and is not particularly motivated for $\Sigma$. Assuming all couplings are natural and $\mathcal{O}(1)$, models L5, SL5, SL6, and S3 cannot significantly modify the electron Yukawa coupling. Consequently, our focus shifts to the phenomenological analysis of other models identified in the previous section. 

\begin{table}[t]
    \centering
    \renewcommand{\arraystretch}{1.4}
    \begin{tabular}{|c|c|c|}
    \hline
        Particle & $\mu \to eee$ & $\mu N\to e N$ \\
        % VLL E
        \hline
        $E\sim (\mathbf{1},\mathbf{1})_{-1}$ & $120\times\sqrt{|[\lambda_E]_2| |[\lambda_E^*]_1|}$\phantom{ai} & $290\times\sqrt{|[\lambda_E]_2| |[\lambda_E^*]_1|}$\phantom{ai} \\
        % VLL Delta1
        \hline
        $\Delta_1\sim (\mathbf{1},\mathbf{2})_{-1/2}$ & $117\times\sqrt{|[\lambda_{\Delta_1}]_2| |[\lambda_{\Delta_1}^*]_1|}$ & $290\times\sqrt{|[\lambda_{\Delta_1}]_2| |[\lambda_{\Delta_1}^*]_1|}$ \\
        % VLL Delta3
        \hline
        $\Delta_3\sim (\mathbf{1},\mathbf{2})_{-3/2}$ & $117\times\sqrt{|[\lambda_{\Delta_3}]_2| |[\lambda_{\Delta_3}^*]_1|}$ & $290\times\sqrt{|[\lambda_{\Delta_3}]_2| |[\lambda_{\Delta_3}^*]_1|}$ \\
        % VLL Sigma1
        \hline
        $\Sigma_1\sim (\mathbf{1},\mathbf{3})_{-1}$ & $84\times\sqrt{|[\lambda_{\Sigma_1}]_2| |[\lambda_{\Sigma_1}^*]_1|}$ & $205\times\sqrt{|[\lambda_{\Sigma_1}]_2| |[\lambda_{\Sigma_1}^*]_1|}$\\ 
        \hline
    \end{tabular}
    \caption{Current constraints from the LFV processes on the vector-like lepton mass measured in TeV. The future measurements will improve the bound on the mass by an order of magnitude.}
    \label{tab:LFV_constraints}
\end{table}

\paragraph{Lepton flavour violation.} 

Vector-like leptons with a generic flavour structure typically induce significant flavour violation, necessitating high mass scales. To maintain a low scale for new physics and allow substantial enhancement of $\kappa_e$, the new physics interactions must preserve the accidental SM lepton flavour symmetries with good accuracy. To illustrate this, consider a scenario where new states couple to first- and second-generation leptons simultaneously. Integrating out vector-like leptons leads to off-diagonal $Z$-boson couplings to charged leptons, described by
\begin{align}
\mathcal{L}_{\rm eff} \supset 
&-\sqrt{g_2^2+g_1^2}\, Z^{\mu} \left( \bar f^i_L \gamma_{\mu}
\left( g_L^{Zf}\delta_{ij}+ \delta g_{L\,ij}^{Zf}\right) f_L^j
+\bar e^i_R \gamma_{\mu}
\left[ g_R^{Ze} \delta_{ij}+ \delta g_{R\,ij}^{Ze}\right] e_R^j
\right)\,,
\label{eq:EffLagZ}
\end{align}
where $f = \nu, e$, and the coupling modifications are:  
\begin{align}
\delta g_{L\,ij}^{Z\nu}=&-\frac{v^2}{2}\left([\cC_{\phi \ell}^{(1)}]_{ij}-[\cC_{\phi \ell}^{(3)}]_{ij}\right)\,,\label{eq:deltagZnuL}\\
\delta g_{L\,ij}^{Ze}=&-\frac{v^2}{2}\left([\cC_{\phi \ell}^{(1)}]_{ij}+[\cC_{\phi \ell}^{(3)}]_{ij}\right)\,,\label{eq:deltagZeL}\\
\delta g_{R\,ij}^{Ze}=&\,-\frac{v^2}{2}[\cC_{\phi e}]_{ij}\,.\label{eq:deltagZeR}
\end{align}  
As shown in Tab.~\ref{tab:treelevelmatching_novarphi}, all models with at least one vector-like lepton induce $Z\mu e$ coupling at tree level, which is strongly constrained by processes such as $\mu \to eee$ and $\mu \to e$ conversion in the presence of a nucleus $N$. In particular, Tab.~\ref{tab:LFV_constraints} summarizes the mass bounds for each vector-like lepton to satisfy these constraints. Current limits include BR$(\mu \to eee) < 1.0 \times 10^{-12}$ at 90\% CL from the SINDRUM spectrometer~\cite{SINDRUM:1987nra} and $\Gamma(\mu^- {\rm Au} \to e^- {\rm Au})/\Gamma_{\rm capture}(\mu^- {\rm Au}) < 7 \times 10^{-13}$ at 90\% CL from the SINDRUM II experiment~\cite{SINDRUMII:2006dvw}, and we use the theory predictions derived in~\cite{Ishiwata:2015cga}. 

For $\mathcal{O}(1)$ couplings to electrons and muons, the new particle mass scale must be $\mathcal{O}(100)$ TeV, limiting their contribution to $\kappa_e$ at observable levels. Moreover, the upcoming experiments aim to improve these bounds significantly. The Mu3e Phase-I experiment is expected to achieve a sensitivity of $2 \times 10^{-15}$ at 90\% CL~\cite{Blondel:2013ia}, while the COMET Phase-I and Mu2e experiments target sensitivities of $7 \times 10^{-15}$ and $8 \times 10^{-17}$ at 90\% CL, respectively, for $\mu \to e$ conversion in aluminum~\cite{COMET:2018auw,Mu2e:2014fns}. These advancements could push the vector-like lepton mass scale to $\mathcal{O}(1000)$ TeV, rendering such scenarios difficult to test even at FCC-ee (see Sec.~\ref{sec:results}).  

Thus, new physics scenarios involving vector-like leptons that enhance the electron Yukawa coupling at observable levels must couple exclusively to electrons. However, even this approach faces stringent constraints due to the chirally enhanced $v/m_e$ contributions to electron dipoles, which we discuss next.

\begin{table}[t]
    \centering
    \begin{tabular}{|c|c|c|c||c|c|c|c|}
    \hline
       Model  & $\eta$ & $|\kappa_e|$ & $\Lambda$ [TeV] & Model & $\eta$ & $|\kappa_e|$ & $\Lambda$ [TeV]\\
        \hline
        L1: $E+\Delta_1$& $1$ & $19$ & $34$ & SL1: $S+E$\phantom{a}& $3$ & $7$ & $59$\\
        \hline
        L2: $E+\Delta_3$& $1/5$ & $92$ & $15$ & SL2: $S+\Delta_1$& $3$ & $7$ & $59$\\
        \hline
        L3: $\Sigma_1+\Delta_3$& $1/5$ & $92$ & $15$ & SL3: $\Xi+\Delta_1$& $-1$ & $17$ & $36$\\
        \hline
        L4: $\Sigma_1+\Delta_1$& $1/9$ & $165$ & $11$ & SL4: $\Xi+\Sigma_1$& $1/3$ & $56$ & $19$\\
        % \hline
        % L5: $\Sigma_1+\Delta_3$& $5$ & $10$ & $46$ & SL5: $\Xi_1+\Delta_3$& $1/2$ & $89$ & $15$\\
        \hline
    \end{tabular}
    \caption{The model dependent coefficient $\eta$ defined in Eq.~\eqref{eq:ae-eta} relating the contributions to $\Delta a_e$ and $\kappa_e$. The corresponding limits on the $\kappa_e$ and the new physics scale $\Lambda$ when the new physics couplings are of $\mathcal{O}(1)$ have been derived requiring $\Delta a_e < 5\cdot 10^{-13}$.}
    \label{tab:(g-2)e}
\end{table}

\paragraph{Chirally enhanced dipole moments.}

Due to the same chiral structure of the operators responsible for modifying the electron Yukawa coupling and the dipole, the same chiral enhancement of the Yukawa coupling will generally be inherited in the electron dipole moments~\cite{Altmannshofer:2015qra, Dermisek:2020cod, Crivellin:2021rbq, Paradisi:2022vqp, Davoudiasl:2023huk}. Focusing on real new physics couplings,\footnote{New Yukawa type interactions can in general feature phases, which could lead to CP-violation strongly constrain by measurements of electric dipole moments. In the spirit of discussing here the maximal allowed Yukawa couplings, we assume that those constrains do not apply, hence no new CP-violation.} the contribution to the anomalous magnetic moment of the electron $a_e = (g_e-2)/2$ can be encoded in terms of the following effective operators of the SMEFT Lagrangian
\begin{equation}
\label{eq:eft-gminus2}
\mathcal{L}_\mathrm{SMEFT} \supset {C_{eB}} \,\big{(}\bar{\ell}_L \sigma^{\mu\nu} e_R\big{)}\phi B_{\mu\nu} + {C_{eW}} \,\big{(}\bar{\ell}_L \sigma^{\mu\nu} e_R\big{)} \sigma^I \phi\, W^I_{\mu\nu} +\mathrm{h.c.}\,.
\end{equation}
The leading new physics contribution, $\Delta a_e$, is given by
\begin{equation}
\Delta a_e \simeq \dfrac{4 m_e v}{\sqrt{2} e } \mathrm{Re}\left(C_{e\gamma}\right)\,,
\end{equation}
where $\smash{C_{e\gamma}= \cos \theta_W\, C_{eB} - \sin \theta_W\, C_{eW}}$. 

\begin{figure}[!t]
    \centering
    \begin{subfigure}[t]{0.3\linewidth}
    \centering
    \begin{tikzpicture}
        \begin{feynman}
            \vertex (a) at (0,0) [dot, red] {}; %[rectangle, draw] {};
            \vertex[right= of a] (b) [dot, green] {};%  [empty dot] {};
            \vertex[above right= of b] (c) [dot, blue] {};
            \vertex[above left= of a] (phi1) {\(\phi\)};
            \vertex[below left= of a] (barlL) {\(\bar{\ell}_L\)};
            \vertex[below right= of b] (eR) {\(\phi\)};
            \vertex[above= of c] (phi2) {\(\phi^*\)};
            \vertex[right= of c] (phistar) {\(e_R\)};
            \vertex[right = of barlL] (gauge1) ;
            \vertex[below left = of eR] (gauge2) ;%{\(W^I_\mu,\, B_\mu\)};
            \diagram*{
                (eR) -- [scalar] (b) -- [fermion, edge label=\(\Psi_1\)] (a) -- [fermion] (barlL);
                (phi1) -- [scalar] (a);
                (b) -- [anti fermion, edge label'=\(\Psi_2\)] (c) -- [scalar] (phi2);
                (c) -- [anti fermion] (phistar);
            };
        \end{feynman}
    \end{tikzpicture}  
    \caption{}
    \label{fig:FD_dipole_a_VLL}
    \end{subfigure}
    \hspace{2cm}
    \begin{subfigure}[t]{0.3\linewidth}
    \centering
    \begin{tikzpicture}
        \begin{feynman}
            \vertex (a) at (0,0) [dot, red] {}; %[rectangle, draw] {};
            \vertex[right= of a] (b) [dot, green] {};%  [empty dot] {};
            \vertex[above right= of b] (c) [dot, blue] {};
            \vertex[above left= of a] (phi1);
            \vertex[below left= of a] (barlL) {\(\bar{\ell}_L\)};
            \vertex[below right= of b] (eR) {\(\phi\)};
            \vertex[above= of c] (phi2);
            \vertex[right= of c] (phistar) {\(e_R\)};
            \diagram*{
            (gauge1) -- [boson, edge label' = \(X_\mu\)] (gauge2);
                (eR) -- [scalar] (b) -- [fermion, edge label=\(\Psi_1\)] (a) -- [fermion] (barlL);
                (b) -- [anti fermion, edge label'=\(\Psi_2\)] (c);
                (c) -- [anti fermion] (phistar);
                (a) -- [scalar, half left, edge label=\(\phi\)] (c);
            };
        \end{feynman}
    \end{tikzpicture}  
    \caption{}
    \label{fig:FD_dipole_b_VLL}
    \end{subfigure}
    \caption{Feynman diagrams involving a pair of vector-like lepton $\Psi_1$ and $\Psi_2$ contributing to the electron Yukawa \textbf{(a)}, and the magnetic moment \textbf{(b)}. Interaction vertices involving the same type of particles are identified by the same colored dot. The gauge boson line assumes insertion wherever possible and $X_\mu$ denotes either $B_\mu$ or $W^I_\mu$.}
    \label{fig:ae_ye_L}
\end{figure}
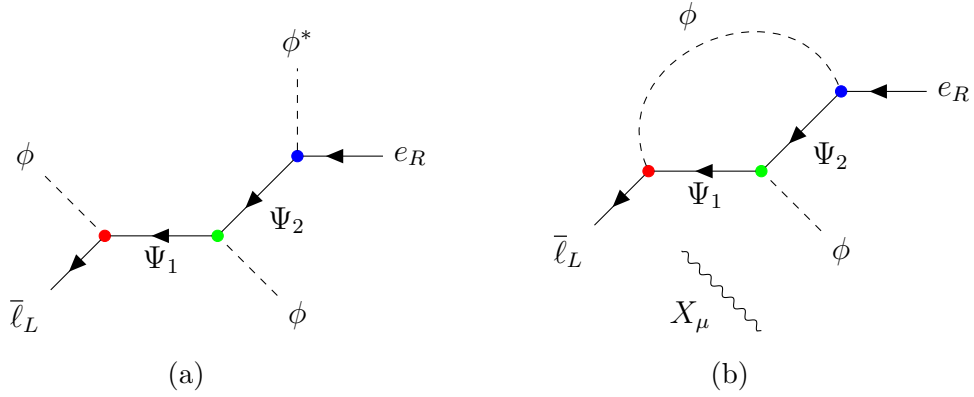

\begin{figure}[!t]
    \centering
    \begin{subfigure}[t]{0.3\linewidth}
    \centering
    \begin{tikzpicture}
        \begin{feynman}
            \vertex (a) at (0,0) [dot, red] {}; %[rectangle, draw] {};
            \vertex[right= of a] (b) [dot, green] {};%  [empty dot] {};
            \vertex[above right= of b] (c) [dot, blue] {};
            \vertex[above left= of a] (phi1) {\(\phi\)};
            \vertex[below left= of a] (barlL) {\(\bar{\ell}_L\)};
            \vertex[below right= of b] (eR) {\(e_R\)};
            \vertex[above= of c] (phi2) {\(\phi\)};
            \vertex[right= of c] (phistar) {\(\phi^*\)};
            \diagram*{
                (eR) -- [fermion] (b) -- [fermion, edge label=\(\Psi\)] (a) -- [fermion] (barlL);
                (phi1) -- [scalar] (a);
                (b) -- [scalar, edge label'=\(\Phi\)] (c) -- [scalar] (phi2);
                (c) -- [scalar] (phistar);
            };
        \end{feynman}
    \end{tikzpicture}  
    \caption{}
    \label{fig:FD_dipole_a}
    \end{subfigure}
    \hfill
    \begin{subfigure}[t]{0.3\linewidth}
    \centering
    \begin{tikzpicture}
        \begin{feynman}
            \vertex (a) at (0,0) [dot, red] {};
            \vertex[right = of a] (b) [dot, green] {};
            \vertex(c) at (0.8,1.) [dot, blue] {};
            \vertex[below left = of a] (barlL) {\(\bar{\ell}_L\)};
            \vertex[below right = of b] (eR) {\(e_R\)};
            \vertex[above right= of c] (phi) {\(\phi\)};
            \vertex[right = of barlL] (gauge1) ;
            \vertex[below left = of eR] (gauge2) ;%{\(W^I_\mu,\, B_\mu\)};
            \diagram*{
                (eR) -- [fermion] (b) -- [fermion, edge label = \(\Psi\)] (a) -- [fermion] (barlL);
                (a) -- [scalar, edge label = \(\phi\)] (c) --  [scalar, edge label = \(\Phi\)] (b);
                (c) -- [scalar] (phi);
                (gauge1) -- [boson, edge label' = \(X_\mu\)] (gauge2);
            };
        \end{feynman}
    \end{tikzpicture}
    \caption{}
    \label{fig:FD_dipole_b}
    \end{subfigure}
    \hfill
    \begin{subfigure}[t]{0.3\linewidth}
    \centering
    \begin{tikzpicture}
        \begin{feynman}
        \vertex (a) at (0,0) [dot, red] {};
        \vertex[right = of a] (b) [dot, purple] {};
        \vertex[below right = of b] (c) [dot, green] {};
        \vertex[above left = of a] (phi) {\(\phi\)};
        \vertex[below left = of a] (barlL) {\(\bar{\ell}_L\)};
        \vertex[below right = of c] (eR) {\(e_R\)};
        \vertex[right= of barlL] (guage1);
        \vertex[below right = of guage1] (guage2);
        \diagram*{
            (eR) -- [fermion] (c) -- [fermion, half right , looseness = 1.1, edge label' = \(\Psi\)] (b) -- [fermion, edge label = \(\Psi\)] (a) -- [fermion] (barlL);
            (a) -- [scalar] (phi);
            (c) -- [scalar, half left, looseness = 1.1, edge label = \(\Phi\)] (b);
            (guage1) -- [boson, edge label' = \(X_\mu\)] (guage2);
        };
        \end{feynman}
    \end{tikzpicture}
    \caption{}
    \label{fig:FD_dipole_c}
    \end{subfigure}
    \caption{Feynman diagrams involving the NP scalar $\Phi$ and the vector-like lepton $\Psi$ contributing to the electron Yukawa \textbf{(a)}, and the magnetic moment \textbf{(b,c)}.}
    \label{fig:FD_dipole}
\end{figure}
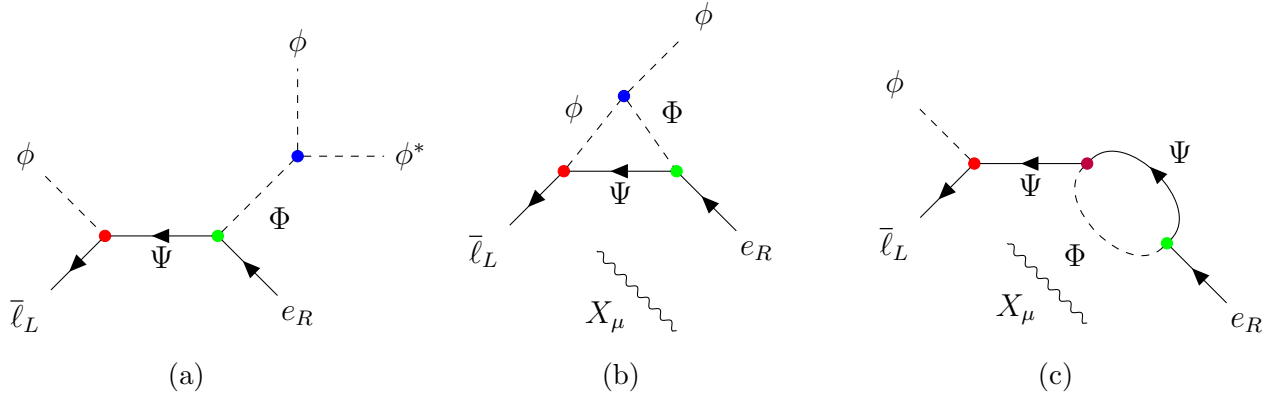

For models involving a pair of vector-like leptons or a vector-like lepton and a scalar, the contribution to $\Delta a_e$ is directly correlated with that to $\kappa_e$. For models with pairs of vector-like leptons, this is illustrated in Fig.~\ref{fig:ae_ye_L}, while for models containing a vector-like lepton in combination with a new scalar, the underlying Feynman diagrams are shown in Fig.~\ref{fig:FD_dipole}. Assuming equal masses for the new physics states, this relationship is
\begin{equation}
\Delta a_e = \eta\, \frac{m_e^2 (\kappa_e -1)}{16\pi^2 v^2}\,,
\label{eq:ae-eta}
\end{equation}
where $\eta$ is a model-dependent rational number. For models where the chiral enhancement in the electron Yukawa coupling is related to the anomalous magnetic moment of the electron, we list the corresponding values of $\eta$ in Tab.~\ref{tab:(g-2)e}. In addition, we list the complementary limit on $\kappa_e$ and the corresponding scale assuming all new physics couplings are of $\mathcal{O}(1)$. We compare the SM prediction~\cite{Czarnecki:1995sz, Aoyama:2012wj,Nomura:2012sb,Giudice:2012ms} with the latest experimental measurement of~$a_e^{\rm exp} = (115\, 965\, 218\, 059 \pm 13) \times 10^{-14}$~\cite{Fan:2022eto} which results in $\Delta a_e = a_e^{\rm exp} - a_e^{\rm SM} = (33.8 \pm 16.1) \times 10^{-14}$ if the most precise value for the fine-structure constant, obtained by measuring the recoil velocity of a rubidium atom absorbing a photon using matter-wave interferometry~\cite{Morel:2020dww}, is used. The heavy states in the models we consider have a negligible impact on extracting the fine-structure constant that involves atomic experiments probing energy scales below the electron mass~\cite{Giudice:2012ms}, thereby allowing $\Delta a_e$ to be their unambiguous test. Finally, we note that models with a vector-like lepton and a scalar, shown in Fig.~\ref{fig:FD_dipole}, violate the relation in Eq.~\eqref{eq:ae-eta} due to extra contributions proportional to the couplings involving new physics states only, as shown in Fig.~\ref{fig:FD_dipole_c}. Such topologies have already been identified in the context of explaining the anomaly in the anomalous magnetic moment of the muon~\cite{Guedes:2022cfy}. Again, we assume a natural scenario without fine-tuned cancellations between this contribution and the one in Eq.~\eqref{eq:ae-eta} (Fig.~\ref{fig:FD_dipole_b}) that would significantly decorrelate $\Delta a_e$ and $\kappa_e$.

\subsubsection{Models with new scalars}
\label{sec:scalars_flavour}

Simplified models involving only scalars that enhance the electron Yukawa coupling necessarily include an additional $SU(2)$ doublet, $\varphi$, which couples to leptons. Excluding the scalar triplet with unit hypercharge, $\Xi_1$, which induces lepton number violation at dimension five, $\varphi$ is the only scalar that generates a flavoured Wilson coefficient in the SMEFT at tree level. Hence, the flavour phenomenology of the scalar extensions is dominated by the Yukawa coupling $[y_\varphi^e]_{ij}$, which encodes the interaction of $\varphi$ with the SM leptons $\ell_L^i$ and $e_R^j$. 

In the heavy new physics limit, the impact of $\varphi$ on low-energy lepton transitions is captured by the four-lepton operator $[\mathcal{O}_{\ell e}]_{ijkl}$ with the matching to the EFT provided in Tab.~\ref{tab:treelevelmatching_varphi}. The contribution to the electron Yukawa coupling arises from the flavour-conserving term $[y_\varphi^e]_{11}$, constrained by its impact on the differential distribution of $e^+e^-$ events in Bhabha scattering, which was precisely measured at LEP~\cite{ALEPH:2013dgf}. The contribution to the differential cross-section due to $[\mathcal{O}_{\ell e}]_{1111}$ reads
\begin{align}
\left(\frac{d\sigma}{d\cos\theta}\right)_{\rm NP} \!\!&= \frac{1}{8 \pi s} 
\left[t^2\left(\frac{e^2}{s} 
+ \frac{(g_2^2+g_1^2)g_{L}^{Ze} g_{R}^{Ze}}{s-m_Z^2}\right)+ (s\leftrightarrow t)\right][\mathcal{C}_{\ell e}]_{1111} \,,
 \label{eq:lep_Bhabba}
\end{align} 
where $t=-s(1-\cos\theta)/2$. The measurements of $d\sigma/d\cos\theta$ were performed for fifteen angular bins in $\cos\theta \in [-0.9, 0.9]$ and seven center-of-mass energies in $\sqrt{s} \in [189, 207]$ GeV. A fit to this data in Ref.~\cite{Falkowski:2015krw} yielded a 95\% CL limit of $[\mathcal{C}_{\ell e}]_{1111}< 1/13.7$ TeV$^{-2}$, which translates to
\begin{equation}
    \frac{M_\varphi}{|[y_\varphi^e]_{11}|}>2.7 \,\,{\rm TeV}\,.
    \label{eq:doublet_constraint}
\end{equation}
We apply this bound to constrain the contribution of simplified scalar-only models to the electron Yukawa coupling. Additionally, stringent lepton flavour violation limits forbid $\varphi$ from having large off-diagonal couplings to leptons. For example, the decay $\mu \to eee$ imposes the following hierarchy~\cite{Falkowski:2015krw}
\begin{equation} 
\frac{|[y_\varphi^e]_{21}|}{|[y_\varphi^e]_{11}|} < 3.4 \cdot 10^{-4}\,,
\end{equation}
though this constraint does not impact the enhancement of $\kappa_e$.

Other scalar states that could modify the electron Yukawa always appear in combination with $\varphi$, such that their couplings could be probed through flavour transitions at one loop level. In particular, the trilinear coupling between two new scalars and the SM doublet, $\kappa_{S\varphi}$ and $\kappa_{\Xi\varphi}$ in Models S1 and S2, respectively, could be probed through the following flavoured Wilson coefficients arising at one loop
\begin{align}
    \lsquare\cC_{\phi l}^{(1\pm3)}\rsquare_{ij} &= \eta_\pm \frac{|\kappa_{\Phi\varphi}|^2 [y_{\varphi}^e]_{ik}[y_{\varphi}^e]_{jk}^*}{16\pi^2 (M_\Phi^2-M_{\varphi}^2)^3}\,\left(M_\Phi^2-M_{\varphi}^2 -(M_\Phi^2+M_{\varphi}^2)\log\left(\frac{M_\Phi}{M_{\varphi}}\right)\right)\,,\label{eq:Cphilpm}\\
    \lsquare\cC_{le}\rsquare_{ijkl} &= \eta \frac{|\kappa_{\Phi\varphi}|^2 [y_{\varphi}^e]_{il}[y_{\varphi}^e]_{jk}^*}{16\pi^2 M_{\varphi}^4}\,\left(1+\log\left(\frac{\mu^2}{M_{\varphi}^2}\right)\right)\,.\label{eq:cleS1S2}
\end{align}
Here, $\Phi=\{S,\Xi\}$ and $\cC_{\phi l}^{(1\pm3)}$ correspond to the Wilson coefficients of the operators obtained by taking the sum or difference of the operators in Eqs.~\eqref{eq:Ophil1} and~\eqref{eq:Ophil3}, directly relevant for the $Z$-coupling modifications in Eqs.~\eqref{eq:deltagZnuL} and~\eqref{eq:deltagZeL}. For Model S1, $\{\eta_+,\eta_-,\eta\}=\{1/2,0,-1/2\}$, and for Model S2, $\{\eta_+,\eta_-,\eta\}=\{1/2,1,-3/2\}$. Additionally, in Model S1 (S2), $[\cC_{\phi e}]_{ij} = -[\cC_{\phi l}^{(1+3)}]_{ij}$ ($[\cC_{\phi e}]_{ij} = -3[\cC_{\phi l}^{(1+3)}]_{ij}$). When reporting Eqs.~\eqref{eq:Cphilpm} and~\eqref{eq:cleS1S2} we have isolated the contributions involving the NP couplings and neglected the ones proportional to the SM Yukawa or gauge couplings. The renormalisation scale $\mu$ in Eq.~\eqref{eq:cleS1S2} originates from the two-point function renormalisation of $\varphi$ induced by dimensionful coupling $\kappa_{\Phi\varphi}$. Hence, great precision in the measurement of the $Z$ couplings to electrons, sensitive to $[\cC_{\phi l}^{(1+3)}]_{11}$ and $[\cC_{\phi e}]_{11}$, at the FCC-ee will allow us to indirectly probe $|\kappa_{\Phi\varphi}|$ which would otherwise be particularly difficult to constrain.

\subsection{Direct searches}
\label{subsec:directsearches}

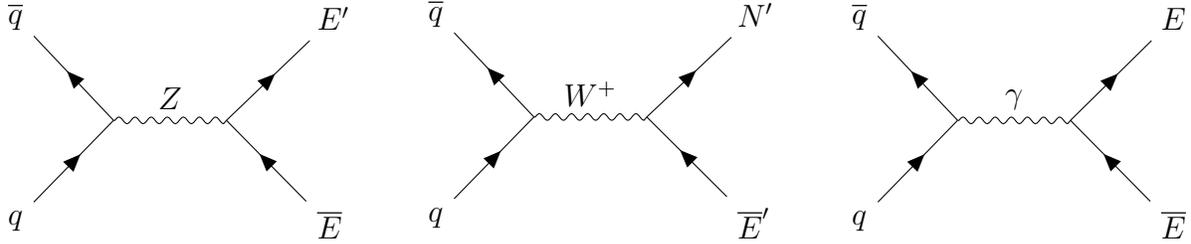
\begin{figure}[t]
    \centering
    \begin{tikzpicture}
        \begin{feynman}
            \vertex (qqZ) at (0,0) ;
            \vertex[above left= of qqZ] (qbar) {\(\bar{q}\)};
            \vertex[below left= of qqZ] (q) {\(q\)};
            \vertex[right= of qqZ] (WEE);
            \vertex[above right= of WEE] (Eprime) {\(E'\)};
            \vertex[below right= of WEE] (E) {\(\bar{E}\)};
            \diagram*{
                (q) -- [fermion] (qqZ) -- [fermion] (qbar);
                (qqZ) -- [boson, edge label = \(Z\)] (WEE);
                (E) -- [fermion] (WEE) -- [fermion] (Eprime);
            };
        \end{feynman}
    \end{tikzpicture}  
    \hspace*{0.5cm}
    \begin{tikzpicture}
        \begin{feynman}
            \vertex (qqW) at (0,0) ;
            \vertex[above left= of qqW] (qbar) {\(\bar{q}\)};
            \vertex[below left= of qqW] (q) {\(q\)};
            \vertex[right= of qqW] (WNE);
            \vertex[above right= of WNE] (Nprime) {\(N'\)};
            \vertex[below right= of WNE] (Eprime) {\(\bar{E}'\)};
            \diagram*{
                (q) -- [fermion] (qqW) -- [fermion] (qbar);
                (qqW) -- [boson, edge label = \(W^+\)] (WNE);
                (Eprime) -- [fermion] (WNE) -- [fermion] (Nprime);
            };
        \end{feynman}
    \end{tikzpicture}
    \hspace*{0.5cm}
    \begin{tikzpicture}
        \begin{feynman}
            \vertex (qqgamma) at (0,0) ;
            \vertex[above left= of qqgamma] (qbar) {\(\bar{q}\)};
            \vertex[below left= of qqgamma] (q) {\(q\)};
            \vertex[right= of qqgamma] (gammaEE);
            \vertex[above right= of gammaEE] (E) {\(E\)};
            \vertex[below right= of gammaEE] (barE) {\(\bar{E}\)};
            \diagram*{
                (q) -- [fermion] (qqgamma) -- [fermion] (qbar);
                (qqgamma) -- [boson, edge label = \(\gamma\)] (gammaEE);
                (barE) -- [fermion] (gammaEE) -- [fermion] (E);
            };
        \end{feynman}
    \end{tikzpicture}  
    \caption{Exemplary diagrams for the production of a pair of VLLs at the LHC in Model L1. The diagram on the left appears after the rotation into the mass basis and shows the production of two VLLs with electric charge $\pm 1$, with $E$ being the singlet and $E'$ the charge component of the doublet $\Delta_1$. The central diagram shows the production of the two $\Delta_1$ components, $N'$ and $E'$, from the decay of the $W$ boson. The diagram on the right showcases the scenario with a photon propagator.}
    \label{fig:VLLproduction_hadroncolliders}
\end{figure}

The ATLAS collaboration has recently reported on direct searches for VLLs coupled to the first and second lepton generations performed at the LHC in Ref.~\cite{ATLAS:2024mrr}. In particular, for the first-generation scenario, the following lower bounds with a $95\%$CL are found
\begin{align}
    M_E &> \SI{320}{\giga\electronvolt}\,,\\
    M_{\Delta_1} &> \SI{1220}{\giga\electronvolt}\,.
\end{align}
However, these bounds are not directly applicable to the scenarios we study. Firstly, the lower bound on $M_E$ is very low, and would not justify the EFT approach. Secondly, only two out of the five VLLs presented in Tab.~\ref{tab:vectorlikeleptons} are considered. Finally, we consider the presence of two VLLs in each model, while Ref.~\cite{ATLAS:2024mrr} considers one VLL per time.

Therefore, we consider the projections for the HL-LHC provided by the CMS collaboration in Sec. 7 of Ref.~\cite{CMS:2024bni} and construct theory predictions for the production cross-section for each of the models containing either one or two VLLs. We consider the masses of the VLL components to be the same, neglecting mass splitting effects; additionally, in the presence of two VLLs, we assume the same mass limit.

VLL production at hadron colliders occurs through electroweak interactions, as illustrated in the Feynman diagrams of Fig.~\ref{fig:VLLproduction_hadroncolliders}. Both neutral and charged gauge bosons are involved, necessitating the inclusion of couplings in the mass basis for all VLL components with the electroweak gauge bosons. The diagram on the left of Fig.~\ref{fig:VLLproduction_hadroncolliders} further demonstrates how, in the broken phase, the mass basis rotation enables interactions between the components of two distinct VLL multiplets and an electroweak gauge boson. For each model considered, the rotation is carried out perturbatively, with full details and results provided in App.~\ref{app:VLL_directsearches}. In Tab.~\ref{tab:vll_massbounds} the projected lower bounds for the VLL masses are presented, having separated the models based on whether there are one or two VLLs. The former case falls into the class of models involving a new scalar and a VLL; for the sake of obtaining lower bounds on the VLL mass, we do not account for the presence of the extra scalar. 

Before moving on to the direct searches for new scalars, we comment on the asterisked value in Tab.~\ref{tab:vll_massbounds}. Considering the extra singlet vector-like lepton, $E$, the lower bound on the mass is found to be approximately $\SI{600}{\giga\electronvolt}$. Given the use of an EFT approach in the next sections, we chose to set the mass to $\SI{1}{\tera\electronvolt}$.
\begin{table}[t]
    \centering
    {\renewcommand{\arraystretch}{1.3}
    \begin{tabular}{|c|c|c|c|c|c|c|c|c|}
    \hline
       Model  &  SL1 & SL2 & SL3 & SL4 & L1 & L2 & L3 & L4 \\\hline
        Mass [TeV] &  $1.0^*$ & 1.7 & 1.7 & 1.8 & 1.8 & 1.9 & 2.2 & 2.1\\\hline
    \end{tabular}
    }
    \caption{HL-LHC projections for direct searches of vector-like leptons, using the more optimistic projections for doublet models reported in~\cite{CMS:2024bni}. We report only the models for which we will perform fits in Sec.~\ref{sec:results}.}
\label{tab:vll_massbounds}
\end{table}

Regarding the direct searches for the models with extra scalars, we note that we assume that there is no direct coupling of the scalar to quarks, which means that gluon fusion production or direct production from initial state quarks is strongly suppressed. The suppression comes with the mixing angle that in an EFT approach is proportional to $\lambda v^2/\Lambda^2$ where $\lambda$ stands generically for $\kappa_S, \kappa_{\Xi}$, or $\lambda_{\varphi}$. At the HL-LHC, direct searches cannot exclude the 2HDM for masses larger than 1 TeV of the heavy Higgs boson and mixing angles below 0.25 assuming that the heavy Higgs boson decays mostly in light Higgs bosons~\cite{Cepeda:2019klc}, implying that $\lambda\lesssim 4$ for an extra heavy scalar with mass at 1 TeV. 
In case the heavy Higgs boson decays mostly to electrons, $Z'$ searches can be reinterpreted, but leading to bounds on the heavy Higgs mass much below the validity of the EFT limit we consider \cite{Davoudiasl:2023huk}. We will hence no further consider direct searches of the new scalars and set the new physics scale for those models to 2 TeV.

%%%%%%%%%%%%%%%%%%%%%%%%%%%%%%%%%%%%%%%%
\subsection{Electroweak precision tests \label{sec:EWPTs}}
%%%%%%%%%%%%%%%%%%%%%%%%%%%%%%%%%%%%%%%%

To analyze the impact of new states on electroweak physics, we construct the electroweak fit using electroweak observables (EWPOs) defined in Tab.~\ref{tab:ewpos}. The corresponding experimental measurements and SM predictions are sourced from \cite{ALEPH:2005ab,ALEPH:2013dgf,Janot:2019oyi,dEnterria:2020cgt,SLD:2000jop,
ParticleDataGroup:2020ssz,CDF:2005bdv,LHCb:2016zpq,ATLAS:2016nqi,D0:1999bqi,ATLAS:2020xea,Breso-Pla:2021qoe}. We define the $\chi^2$-function as
\begin{equation}
\chi_{\rm EWPO}^2 = \sum_{ij}[O_{i,{\rm exp}}-O_{i,{\rm th}}] (\sigma^{-2})_{ij}[O_{j,{\rm exp}}-O_{j,{\rm th}}] \,,
\label{eq:chi2_EWPO}
\end{equation}
with $\sigma^{-2}$ being the inverse of the covariance matrix~\cite{ALEPH:2005ab,ALEPH:2013dgf}. Moreover, we work in the $\{\alpha_{EM},m_Z,G_F\}$ input scheme such that the relevant effective Lagrangian describing interactions of the electroweak gauge bosons with leptons reads
\begin{align}
\mathcal{L}_{\rm eff} &\supset - g_2 \left[\left(W^{+\mu} j^{-}_{\mu}+{\rm h.c.}\right)+ Z^\mu j_\mu^Z\right]+\frac{g_2^2 v^2}{4}(1+\delta m_W)^2 W^{+\mu} W^-_{\mu}+\frac{g_2^2 v^2}{8 c_W^2}Z^{\mu} Z_{\mu}\,,\\
j^{-}_{\mu} &= \frac{1}{\sqrt{2}}\bar \nu_L^i \gamma_{\mu} \left(\delta_{ij}+\delta g_{ij}^{W \ell }\right)e_L^j\,,\\
j_\mu^Z &= \,\frac{1}{c_W} \left[\bar f^i_L \gamma_{\mu}
\left( g_L^{Zf}\delta_{ij}+ \delta g_{L\,ij}^{Zf}\right) f_L^j
+\bar e^i_R \gamma_{\mu}
\left( s_W^2 \delta_{ij}+ \delta g_{R\,ij}^{Ze}\right) e_R^j
\right]\,,
\label{eq:EW_eff_Lag}
\end{align}
where $f=\nu,e$, and
\begin{align}
g_L^{Z\nu}=\frac{1}{2}\,,\quad\quad g_L^{Ze}=-\frac{1}{2}+s_W^2\,,
\end{align}
with $c_W$ ($s_W$) being the cosine (sine) of the Weinberg angle. The $Z$-boson coupling modifiers $[\delta g^{Ze}_{L,R}]_{ij}$ are defined in Eqs.~\eqref{eq:deltagZnuL}--\eqref{eq:deltagZeR}, while the $W$-coupling modifier to leptons is
\begin{equation}
    \delta g^{W\ell}_{ij}=\delta g_{L\,ij}^{Z\nu}-\delta g_{L\, ij}^{Ze} = v^2 \lsquare \cC_{\phi l}^{(3)}\rsquare_{ij}\,.
\label{eq:deltaWL}
\end{equation}
Finally, the $W$ mass modification reads
\begin{align}
\delta m_W=
\frac{v^2 g_1^2}{4(g_2^2-g_1^2)}\left([\cC_{\ell\ell}]_{1221}-2[\cC^{(3)}_{\phi\ell}]_{22}-2[\cC^{(3)}_{\phi\ell}]_{11}\right)-\frac{v^2 g_2^2}{4(g_2^2-g_1^2)}\cC_{\phi D}
-\frac{v^2 g_2 g_1}{g_2^2-g_1^2}\cC_{\phi WB}\,.\label{eq:Wmass_mod}
\end{align}

\begin{table}[t]
    \centering
    \renewcommand{\arraystretch}{1.5}
    \begin{tabular}{|c|c||c|c|}
        \hline
          Observable & Definition & Observable & Definition \\\hline
          $\Gamma_Z$ & $\sum_f \Gamma(Z\to f\bar f)$ & $R_{uc}$ & $\frac{\Gamma(Z\to u\bar u) + \Gamma(Z\to c\bar c)}{2\sum_q \Gamma(Z\to q\bar q)}$ \\ \hline
          $\sigma_{\rm had}$ & $\frac{12\pi}{m_Z}\frac{\Gamma(Z\to e^+ e^-)\Gamma(Z\to q\bar q)}{\Gamma_Z^2}$ & $m_W$ & $m_W$ \\ \hline
          $R_f$ & $\frac{\Gamma(Z\to f\bar f)}{\sum_q \Gamma(Z\to q\bar q)}$ & $\Gamma_W$ & $\sum_{f_1, f_2} \Gamma(W\to f_1 f_2)$ \\ \hline
          $A_f$ & $\frac{\Gamma(Z\to f_L \bar f_L)-\Gamma(Z\to f_R \bar f_R)}{\Gamma(Z\to f\bar f)}$ & ${\rm BR}(W \to \ell\nu)$ & $\frac{\Gamma(W \to \ell\nu)}{\Gamma_W}$  \\ \hline
          $A_{\rm FB}^{0,\ell}$ & $\frac{3}{4} A_e A_\ell$  &$R_{W_c}$ & $\frac{\Gamma(W\to cs)}{\Gamma(W\to ud) + \Gamma(W\to cs)}$    \\ \hline
          $A_c^{\rm FB}$ & $\frac{3}{4} A_e A_c$  &  $A_b^{\rm FB}$  & $\frac{3}{4} A_e A_b$\\ \hline
    \end{tabular}
    \caption{EWPOs and their definitions used to construct the EW fit. Their explicit expressions in SMEFT are taken from~\cite{Allwicher:2023aql}.}
    \label{tab:ewpos}
\end{table}

\paragraph{Models with vector-like leptons.} 
The tree-level operators defined in Eqs.~\eqref{eq:Ophil1}--\eqref{eq:Ophie}, which modify electroweak gauge boson couplings to leptons, involve at most two new physics couplings, as summarized in Tab.~\ref{tab:treelevelmatching_novarphi}. Additionally, universal one-loop contributions arise through box diagrams involving heavy vector-like leptons
\begin{align}
\delta g_{L\,ii}^{Zf}=
-\frac{v^2}{4}\left(T_f^3+\frac{g_1^2}{g_2^2-g_1^2}Q_f\right)\cC_{\phi D}-v^2\frac{g_2g_1}{g_2^2-g_1^2}Q_f\,\cC_{\phi WB}\,,
\label{eq:universal_shift}
\end{align}
where $T^3_f$ and $Q_f$ denote the weak isospin and electric charge of the fermion $f$, respectively. Therefore, in Models L1-L5, these contributions make the electroweak fit sensitive to the Yukawa coupling between two vector-like leptons and the Higgs boson which enters $\kappa_e$. This is essential to probe all three couplings that determine the contributions to $\cC_{e\phi}$ constraining the possible size of the electron Yukawa couplings through the electroweak precision measurements.

\paragraph{Models with scalars.} 

The scalar states that are triplet under $SU(2)$
impact electroweak physics through tree-level contributions to $\cO_{\phi D}$ presented in Tab.~\ref{tab:treelevelmatching_novarphi}. From Eqs.~\eqref{eq:Wmass_mod} and~\eqref{eq:universal_shift}, this would lead to modifications in the $W$-boson mass and a universal shift of the lepton couplings to electroweak gauge bosons. Furthermore, the couplings of $\varphi$ necessary to induce modifications in the electron Yukawa in all models with scalars, enter the electroweak fit at one loop through modifications of the $Z$ and $W$ couplings to leptons. The explicit expressions can be recovered from Eq.~\eqref{eq:Cphilpm}. Finally, Models S1 and S2 match to $\cC_{\phi D}$ as follows
\begin{align}
    \cC_{\phi D}^{(\rm S1)} &= 
    \frac{g_1^2 |\kappa _{\text{s$\varphi $}}|^2}{480 \pi ^2 \Lambda ^4}\,,\\
    \cC_{\phi D}^{(\rm S2)} &= \frac{\kappa _{\Xi } \kappa _{\Xi \varphi }^* \lambda _{\varphi }^* \Big[5+4 \log (\frac{\mu ^2}{\Lambda ^2})\Big]}{8 \pi ^2 \Lambda ^4}+\frac{|\kappa _{\Xi }|^4 \Big[5-2 \log(\frac{\mu ^2}{\Lambda ^2})\Big]-8 |\kappa _{\Xi }|^2 |\kappa _{\Xi \varphi }|^2 \Big[4+5 \log (\frac{\mu ^2}{\Lambda ^2})\Big]}{16 \pi ^2 \Lambda ^6}\,,
\end{align}
where we have kept only the leading dependence on the NP couplings and set the mass of the heavy scalars to be the same and equal to $\Lambda$. The full expressions in the equal mass limit can be found in the ancillary notebook attached to this work.

%%%%%%%%%%%%%%%%%%%%%%%%
\subsection{Higgs physics}
\label{subsec:HiggsPhysics}
%%%%%%%%%%%%%%%%%%%%%%%%
\begin{table}
\centering
{\renewcommand{\arraystretch}{1.2}
\begin{tabular}{|c|c|c|}
    \hline
    Decay channels &  Expected uncertainties $\left[\%\right]$\\
    \hline
    $h\rightarrow$ any     & $\pm0.5$  \\
    $h\rightarrow b\bar{b}$    & $\pm0.3$ \\
    $h\rightarrow W^+W^-$      & $\pm1.2$ \\
    $h\rightarrow ZZ$          & $\pm4.4$  \\
    $h\rightarrow\tau^+\tau^-$ & $\pm0.9$ \\
    $h\rightarrow\gamma\gamma$ & $\pm9.0$ \\
    $h\rightarrow\mu^+\mu^-$   & $\pm19$  \\
    \hline
\end{tabular}
}
\caption{Expected uncertainties for $\sigma_{Zh}$ times Higgs branching ratios in different channels at $\SI{240}{\giga\electronvolt}$ FCC-ee run~\cite{Bernardi:2022hny}, used to construct the Higgs physics fit.}
\label{tab:higgsfit_fcc_obs}
\end{table}
We will now consider constraints from higher energy runs of the FCC-ee. In particular, the programmed run of FCC-ee at $\SI{240}{\giga\electronvolt}$ optimises the associated production of a $Z$ and a Higgs boson, allowing for a combined fit involving both projected electroweak and Higgs physics observables. In the following, we outline the Higgs fit procedure and the relevant observables. 

\paragraph{Higgstrahlung cross-section.} 
We construct the NLO cross-section for the associated $Zh$ production ($\sigma_{Zh}$) using App.~B of Ref.~\cite{Asteriadis:2024xts}. In femtobarns, its expression reads
\begin{align}
    \sigma^{\rm SMEFT, NLO}_{Zh} = \sigma_{Zh}^{\rm SM, NLO} +  &\left(27.8 \,\cC_{\phi B} + 1.47 \,\cC_{\phi D} + 23.3\, \cC_{\phi \Box} + 100\, \cC_{\phi W} +\right.\nonumber\\
                            &  + 53.5\, \cC_{\phi WB} -158\,\lsquare\cC_{\phi e}\rsquare_{11} + 170\,\lsquare\cC^{(1)}_{\phi \ell}\rsquare_{11} + 190\,\lsquare\cC^{(3)}_{\phi \ell}\rsquare_{11} +\nonumber\\
                            &\left.-1.67\, \cC_\phi + 1.85\cdot10^{-3}\,\lsquare\cC_{le}\rsquare_{1111}\right)\SI{}{\tera\electronvolt}^2\, {\rm fb}\,, 
\label{eq:sigmaZh_NLO}\end{align}
with $\sigma_{Zh}^{\rm SM, NLO}=\SI{194}{\femto\barn}$. We have cross-checked this result against the online database for \texttt{SMEFiT}~\cite{Celada:2024mcf} and found it to be in agreement\footnote{With the exception of the contribution of the operator $\cO_{\ell e}$, which is not included in the database. }. The Wilson coefficients in the third line enter the cross-section only at NLO. However, from the point of view of our models, these two operators are of interest for the models involving NP scalars and we include them. Indeed, $\varphi$ generates both operators at the tree level (see Tab.~\ref{tab:treelevelmatching_novarphi}). As for the other two scalar particles, $S$ and $\Xi$ from Tab.~\ref{tab:treelevelmatching_varphi}, one can appreciate that only $\cO_\phi$ is tree-level generated. Given the large number of couplings introduced by the models with scalars, we chose to set to the best-fit value all the couplings that do not enter the tree-level expressions for $\cC_{e\phi}$ (Tab.~\ref{tab:Cephi_scalarpairs}).
% \par In a first step, we use Eq.~\eqref{eq:sigmaZh_NLO} to write down the constraint from the Higgs cross-section measurement that is added to the EW $\chi^2$-function. Following Refs.~\cite{deBlas:2022ofj, Bernardi:2022hny} (first row of Tab.~\ref{tab:higgsfit_fcc_obs}) and considering the accuracy of $0.5\%$ on the measurement of the total cross-section at center of mass energy $\SI{240}{\giga\electronvolt}$, the contribution to the total $\chi^2$ is given by
% \begin{equation}
%     \chi^2_{\rm Higgs, \,1} = \left(\frac{\sigma_{Zh}^{\rm SMEFT, LO} - \sigma_{Zh}^{\rm SM}}{\sigma_{Zh}^{\rm SM}}\right)^2 \cdot 0.005^{-2}\,.
% \label{eq:Higgsfit_1}\end{equation}
% \NS{Should it be $\sigma_{Zh}^{\rm SMEFT, NLO}$?}

\paragraph{Signal strengths.}
Next, one can make use of the individual measurements of the different Higgs final states shown in Tab.~\ref{tab:higgsfit_fcc_obs}. It contains the projected uncertainties on the product between the $Zh$ production cross section times the branching ratio for the various Higgs decay channels.
We consider the Higgs decays into fermions such as the $b$ quarks, $\tau$-leptons and muons, and vector bosons such as $\gamma$, $W$ and $Z$.  For a given decay channel $i$, the signal strength is defined as 
\begin{equation}
    \mu^{\rm theo}_i = \frac{\sigma_{Zh}^{\rm SMEFT}\cdot \text{Br}^{\rm SMEFT}_{\small h\rightarrow i}}{\sigma_{Zh}^{\rm SM}\cdot \text{Br}^{\rm SM}_{\small h\rightarrow i}} = \frac{\sigma_{Zh}^{\rm SMEFT}}{\sigma_{Zh}^{\rm SM}} \cdot \frac{\Gamma^{\rm SMEFT}_{\small h\rightarrow i}}{\Gamma^{\rm SM}_{\small h\rightarrow i}} \cdot \frac{\Gamma_h^{\rm SM}}{\Gamma_h^{\rm SMEFT}}\,. 
    \label{eq:signalstrength}
\end{equation}
Considering the last expression, the first ratio is known from Eq.~\eqref{eq:sigmaZh_NLO}, whereas the remaining two can be adapted from Ref.~\cite{Brivio:2019myy}. In particular, we drop the $U(3)^5$ flavour symmetric limit of Ref.~\cite{Brivio:2019myy} and impose that the NP is dominantly coupled to electrons as required by flavour physics constraints. For illustration, we report the explicit expression for the Higgs decay width in the SMEFT, accounting also for an enhanced electron Yukawa coupling:
\begin{align}
    \Gamma_h^{\rm SMEFT} =&\,\,\Gamma_h^{\rm SM} \left[1 - 1.50v^2 \cC_{\phi W} -1.21 v^2 \cC_{\phi W}+ 1.21 v^2 \cC_{\phi WB}  + 1.83 v^2 \cC_{\phi\Box} - 0.43 v^2 \cC_{\phi D } \right. \nonumber\\
    & \left.\quad\quad\quad \,-\, 0.0003 v^2 \cC_{\phi l}^{(1)}  -0.773 v^2\cC_{\phi l}^{(3)} - 0.0002 v^2\cC_{\phi e}\right.\\
    & \left. +2v^2 \left(\text{BR}^{\rm SM}(h\rightarrow\gamma\gamma) + \text{BR}^{\rm SM}(h\rightarrow gg) \right) \cC_{\phi,{\rm kin}}\right] + \Gamma^{\rm r}_e \kappa^2_e   \nonumber\,. \label{eq:higgsdecaywidthsmeft} 
\end{align}
The first two rows are reported in Ref.~\cite{Brivio:2019myy}, having already dropped the coefficients that are always zero for the models considered, i.e., those associated with operators involving quarks or gluons. The third row was added to account for the rescaling by $\cC_{\phi, \rm kin}$ (see Eq.~\eqref{eq:C_kin}) of the $h\rightarrow\gamma\gamma$, $h\rightarrow gg$ widths. The enhancement of the $h\rightarrow e\bar{e}$ decay channel is considered in the last row. The decay width $\Gamma^{\rm r}_e$ is obtained by rescaling the $h\rightarrow \mu^+\mu^-$ width obtained from \cite{HBRs} by $m_e^2/m_\mu^2$ \cite{ParticleDataGroup:2024cfk}\footnote{We would like to take note, that this approach is only valid at tree-level, but for small couplings higher order diagrams in which the Higgs does not couple directly to the electron would dominate. Since this is the case for small $\kappa_e$, the contribution of this term to the total width becomes negligible so the mistake we make remains negligible.}, finding
\begin{equation}
    \Gamma_\mu^{\rm SM} = 8.9784\cdot 10^{-7} \, \text{GeV}\, \Rightarrow \, \Gamma_e^{\rm r} = 2.100\cdot 10^{-11}\,\text{GeV}\,.
\end{equation}
Given the involved form of Eq.~\eqref{eq:signalstrength}, for each channel we perform an expansion in $v^2/\Lambda^2$ to linearise in Wilson coefficients the SMEFT predictions and stay strictly within the expansion of the effective field theory. As an example, the signal strength for the decay channel into bottom quarks reads
\begin{align}
    \mu_{b\bar{b}} = 1 +v^2 &\left(3.86 \,\cC_{\phi B} + 2.15 \,\cC_{\phi\Box} + 5.49\cdot 10^{-2}\, \cC_{\phi D} - 
 13.5 \,\lsquare\cC_{\phi e}\rsquare_{11}+ 14.4 \,\lsquare\cC_{\phi\ell}^{(1)}\rsquare_{11} + \right.\nonumber\\
 & \quad + 16.2 \,\lsquare\cC_{\phi\ell}^{(3)} \rsquare_{11} + 9.71\, \cC_{\phi W} + 3.33 \,\cC_{\phi WB} + 3.46\cdot10^{-3} \,\cC_{e\phi} + \nonumber \\
 & \quad\left. + 0.142\, \cC_\phi + 1.57\cdot 10^{-3}\,\lsquare\cC_{\ell e}\rsquare_{1111}\right)\,.
\label{eq:mubb_signalstrength}
\end{align}
The total $\chi^2$ function is then
\begin{equation}
    \chi^2_{\rm Higgs} = \sum_{i, j} \left(\mu_i^{\rm  SMEFT} - 1\right)\, \sigma^{-2}_{ij}\, \left(\mu_j^{\rm  SMEFT} - 1\right)\,,
\label{eq:Higgsfit_2}\end{equation}
where $i, j$ identify the available decay channels and the SM prediction is $\mu_i^{\rm SM}=1$. The inverse of the correlation matrix is obtained as 
\begin{equation}
    \sigma^{-2} = \left(X^{\rm T} \cdot \mathbb{1} \cdot X\right)^{-1}\,,
\label{eq:sigma-2_Higgs2}
\end{equation}
where $X$ is used to denote the uncertainties shown in Tab.~\ref{tab:higgsfit_fcc_obs}. Due to their unavailability, we do not include the experimental correlations between the different signal strengths.
% We do not include the correlations between the different signal strengths. However, given the common presence of the $Zh$ production cross section, it is natural to expect a correlation. \NS{Write something here on the availability of this...}
%%%%%%%%%%%%%%%%
\section{Results}
\label{sec:results}
%%%%%%%%%%%%%%%%
Having introduced all the relevant constraints for our study, we will now apply them to the different models. Commonly to all of them, we define the $\chi^2$-distribution which is a function of NP parameters and incorporates the constraints discussed above. Finally, we report $\kappa_e$ in each model obtained by maximisation subject to the condition
\begin{equation}
    \chi^2 <  \chi^2\Big|_{\rm min}\!\! + \chi^2(n,95\%)\,.
    \label{eq:chi2_condition}
\end{equation}
Here, $\chi^2\Big|_{\rm min}$ is the $\chi^2$ minimum, while $\chi^2(n,95\%)$ is the value of the $\chi^2-$distribution with $n$ degrees of freedom and $p$-value of 0.05 corresponding to the $95\%$ CL interval. We present the details regarding different models below.
\begin{table}[]
    \centering
    {\renewcommand{\arraystretch}{1.7}
    \begin{tabular}{|c|c|c|c|}
    \hline
    Particle content    &  $\varphi$   &   $  \varphi+S$   &   $\varphi+\Xi$   \\  
    \hline
    $\kappa_e$          &  266         & 925               &      323       \\
    \hline                            
    \end{tabular}
    }
    \caption{Maximal allowed values for the electron Yukawa coupling enhancement at a $95\%$ CL for the models with scalars. From left to right, these are the models with a scalar doublet extension, and Models S1 and S2. }
    \label{tab:newke_scalars}
\end{table}
\paragraph{Models with scalars.}  Let us start from the models with new scalars defined in Sec.~\ref{subsec:models_scalars}, which we consider to have mass $M_\Phi= \SI{2}{\tera\electronvolt}$, safely above the current and future limits from direct searches (see Sec.~\ref{subsec:directsearches}). As anticipated, lepton number violation excludes Model S3 from inducing large $\kappa_e$ (see Sec.~\ref{subsec:flavourphysics}), leaving only the SM extension with the extra scalar doublet $\varphi$ and Models S1 and S2 to be studied. The latter two involve two NP scalars: $\varphi$ and, respectively, the singlet $S$ and the $SU(2)$ triplet $\Xi$. The fit results are summarised in Tab.~\ref{tab:newke_scalars} and we comment on them below.

The state $\varphi$ is the only single mediator extension of the SM capable of significant electron Yukawa modification. In this case, the $\chi^2$-distribution is constructed as follows
\begin{equation}
    \chi^2_{\varphi} = \chi^2_{\rm EWPO}\lround y^e_\varphi\rround + \chi^2_{\text{Higgs}}\lround y^e_\varphi, \lambda_\varphi\rround + 3.84\lround 17.8 \,\frac{\SI{2.7}{\tera\electronvolt}}{M_\varphi}\, \lsquare y^e_\varphi\rsquare_{11} \rround^2  \,,
    \label{eq:chi2_scalar}
\end{equation}
where the coupling dependencies have been explicitly written out to identify the origin of the constraints for $\lambda_\varphi$ and $y^e_\varphi$. 
% If $i=1$, then the expression from Eq.~\eqref{eq:Higgsfit_1} is employed, else if $i=2$ we are using Eq.~\eqref{eq:Higgsfit_2}.

The third term invokes the constraints on $y^e_\varphi$ imposed by Bhabha scattering derived in Eq.~\eqref{eq:doublet_constraint}. The further factor of 17.8 represents the optimistic FCC-ee projection assuming no theoretical and systematic uncertainties. Given this, the limits would be dominated by the statistics, and repeating the LEP runs (see Sec.~\ref{sec:scalars_flavour}) at the FCC-ee would result in the bound on the NP scale which is approximately $10^{5/4}\simeq 17.8$ times larger. The scaling comes from the fact that $N_{\rm FCC-ee} = 10^5 N_{\rm LEP}$, with $N_{\rm FCC-ee}$ and $N_{\rm LEP}$ being the total number of observed events at FCC-ee and LEP, and the fact that NP effects are encoded in $d=6$ SMEFT operators. Such term is multiplied by 3.84, associated with one degree of freedom in the $\chi^2$-distribution for a constraint derived at a $95\%$ CL.

Overall, we find that the largest allowed enhancement for the electron Yukawa coupling with a $95\%$ CL is $\kappa_e^{\rm max} = 265$. Such value is undoubtedly above both the projected sensitivity for the HL-LHC of $|y_e|<120\, y_e^{\rm SM}$~\cite{Cepeda:2019klc} and the expected bound from two years of running of the FCC-ee, which is $|y_e|<1.6\, y_e^{\rm SM}$~\cite{dEnterria:2021xij}. Such value is though comparable with the upper limit of $y_e<260\, y_e^{\rm SM}$ determined by the ATLAS collaboration in Ref~\cite{ATLAS:2019old}. The quartic coupling, $\lambda_\varphi$, can only be loosely constrained by Higgs physics. In Tab.~\ref{tab:lambdavarphi_intervals} we present the allowed intervals at a 95\% CL of $\lambda_\varphi$ and find it to be in agreement with the interval of $|\lambda_\varphi|\lesssim 7.6$ found by the  \texttt{SMEFiT}~\cite{Giani:2023gfq,terHoeve:2023pvs,Celada:2024mcf} collaboration. In the same table we also include the allowed intervals for $\lambda_\varphi$ for the models with two NP scalars as a comparison. 

Moving on to the models with two scalars, two assumptions are made. First of all, we set the couplings in the scalar potential that do not enter $\cC_{e\phi}$ (tree-level matching results in Tab.~\ref{tab:Cephi_scalarpairs}) to their best-fit value. For Models S1 and S2, this means four couplings remain to be constrained. Secondly, the two dimensionful couplings ($\kappa_\Phi$ and $\kappa_{\Phi\varphi}$ with $\Phi=S, \,\Xi$) are set to be the same; effectively, this reduces the number of free parameters to three. 

% The first assumption implies that both the Higgs Physics set of constraints have to be adapted before performing the fits. Indeed, comparing the tree-level matching results for $\cO_{e \phi }$ in Tab.~\ref{tab:Cephi_scalarpairs} and those for the operator $\cO_\phi$ in Tab.~\ref{tab:treelevelmatching_novarphi}, neither $S$ nor $\Xi$ can contribute to $\cO_\phi$ after the simplification. \BE{adapted this part from what was originally at the end of section 3.4} 
In addition to $\cO_{e\phi}$, models with two scalars always give a contribution to $\cO_{\phi}$ (in Eq.~\eqref{eq:Ophi}) at tree level. This operator affects the trilinear Higgs self-coupling, $\lambda_{hhh}$, which is notoriously difficult to measure, with the HL-LHC projections in Higgs pair production leading to $\lambda_{hhh}/\lambda_{hhh}^{SM}=1-2 \frac{v^4}{m_h^2}C_{\phi}+3 v^2 (C_{\phi\Box}-1/4 C_{\phi D})\in [0.5, 1.6]$ at 68\% CL \cite{ATLAS:2022hsp}. 
At the FCC-ee in the $\SI{240}{\giga\electronvolt}$ and $\SI{365}{\giga\electronvolt}$ runs the trilinear Higgs self-coupling can be measured indirectly by electroweak corrections \cite{McCullough:2013rea}, and in combination with the HL-LHC, it can be constrained to $\lambda_{hhh}/\lambda_{hhh}^{SM}\in [0.57, 1.44]$ \cite{DiVita:2017vrr}. In our study, the FCC-ee constraints on the trilinear self-coupling are encoded in Higgs physics constraints since the operator $\cO_\phi$ affects the $Zh$ production cross section at NLO, see Eq.~\eqref{eq:sigmaZh_NLO}. 

For Model S1 which involves the doublet field $\varphi$ and the singlet $S$, a combination of the couplings in the scalar potential could lead to cancellations in the trilinear Higgs self-coupling. Therefore, being aware of this possible flat direction, we do not take into account constraints from the Higgs trilinear coupling in the combined fit for Model S1. Instead, for Model S2, involving the doublet $\varphi$ and the triplet $\Xi$, the trilinear Higgs modification due to the triplet state is proportional to $\kappa_{\Xi}^2/M_{\Xi}^2$ which is small due to its tree-level contribution to the $W$ boson mass modification. Hence in this scenario we do not expect such a cancellation to occur and consider all contributions to $\cC_\phi$ (see Tab.~\ref{tab:treelevelmatching_varphi}).

Overall, we find an enhancement of $\kappa^{\rm max}_e$ of to $925$ in Model S1 (see Tab.~\ref{tab:newke_scalars}). Instead, Model S2 leads to approximately the same results as the single mediator model. This can be understood as the triplet contribution is rather suppressed with respect to the contribution of $\varphi$ since it leads to a tree-level contribution in $C_{\phi D}$ strongly constrained by electroweak precision tests. Hence, its contribution remains nearly negligible, and the results we find closely resemble the ones in the single mediator model. 

\begin{table}[]
    \centering
    {\renewcommand{\arraystretch}{1.7}
    \begin{tabular}{|c|c|c|c|}
    \hline
    Model    & $\varphi$  & $\varphi + S$ &   $\varphi + \Xi$  \\
    \hline                          
    $\lambda_\varphi$        &   $\lsquare -1.57; 1.57\rsquare$           &    $\lsquare-5.37;4.80\rsquare$            &      $\lsquare-1.67;1.67\rsquare$     \\                            
    \hline
    \end{tabular}
    }
    \caption{Allowed intervals for quartic coupling $\lambda_\varphi$ of the second Higgs doublet, referring to a 95\% CL. }
    %These values can be compared to the one found by the \texttt{SMEFiT}~\cite{Giani:2023gfq} collaboration: considering constraints from LHC Higgs data, they find a bound of $|\lambda_\varphi|\lesssim 7.6$ \cite{terHoeve:2023pvs,Celada:2024mcf}. \NS{Two rows originate from two different Higgs fits? Also SMEFiT does not constrain $\lambda_\varphi$ in the scenarios $\varphi+S$ and $\varphi+\Xi$? - maybe the second sentence is unnecessary.}}
    \label{tab:lambdavarphi_intervals}
\end{table}

The rather large allowed values for $\kappa_e$ in the models with scalars could also be understood as the best probes of the coupling $\lambda_\varphi$. Indeed, even the current precision on the electron Yukawa results in a better limit than the one obtained from the Higgs observables introduced in Sec.~\ref{subsec:HiggsPhysics}. 

\begin{figure}
    \centering
    \includegraphics[width=0.7\linewidth]{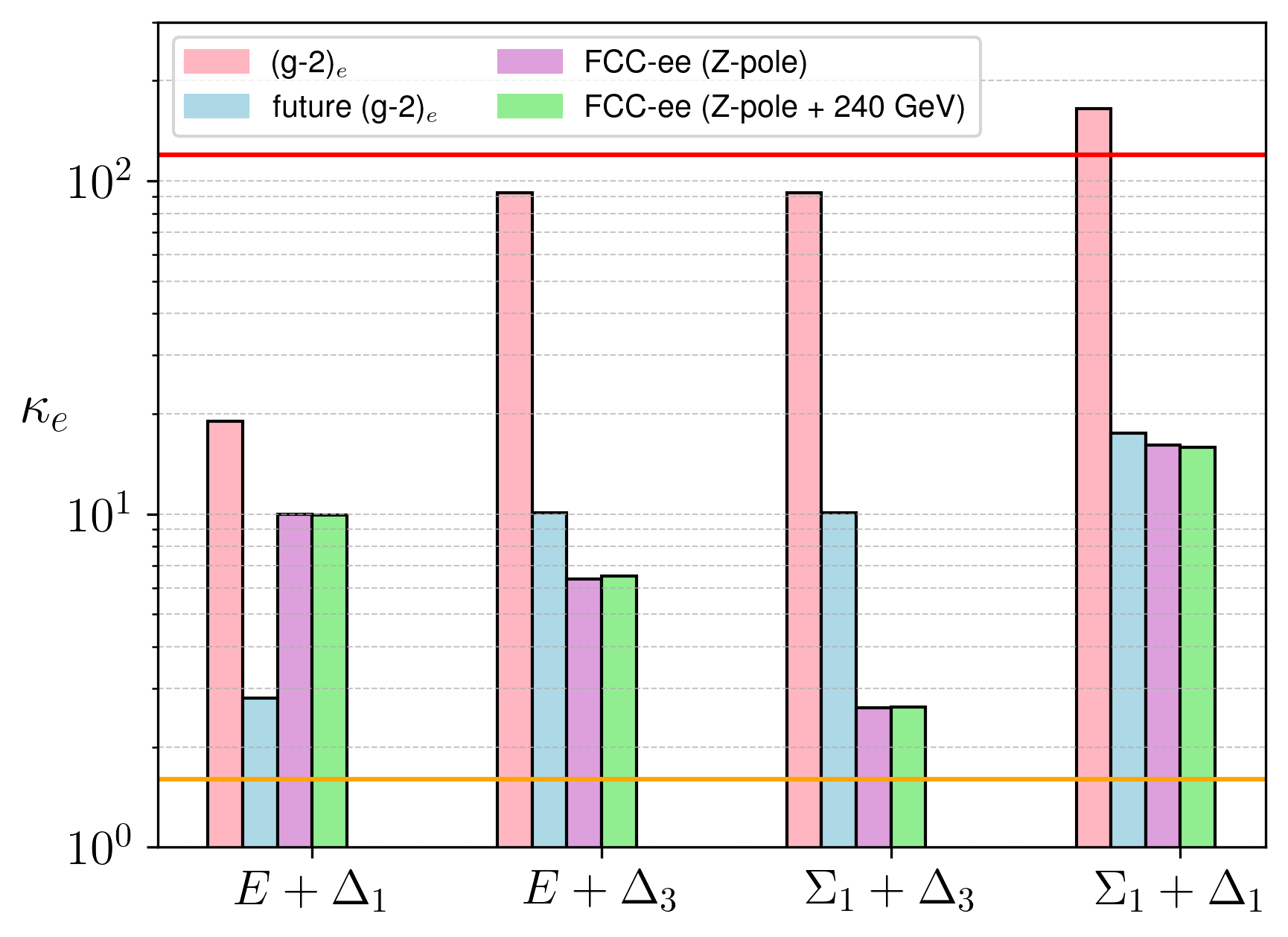}
    \caption{Final plot showing our results for the maximal allowed value for the coupling enhancement $\kappa_e$ with a 95\% CL.}
    \label{fig:newke_vllpairs}
\end{figure}

\paragraph{Models with vector-like lepton pairs. }
We now turn to the models involving pairs of vector-like leptons, as defined in Sec.~\ref{subsec:models_vllpairs}. Fig.~\ref{fig:newke_vllpairs} summarizes the maximum allowed values of the coupling modifier at the $95\%$ confidence level. The figure presents two horizontal lines associated with the HL-LHC projected sensitivity $y_e< 120\, y_e^{\rm SM}$ (red) and with the expected limit of $|y_e|<1.6\, y_e^{\rm SM}$ (orange) which could be reached after two years of a dedicated run at the FCC-ee. The values for the masses of the new states are fixed to those reported in Tab.~\ref{tab:vll_massbounds} coming from the HL-LHC direct search limits.

The red bars show the current impact of the electron $g-2$ in constraining the possible size of its Yukawa coupling. It reflects the entries in Tab.~\ref{tab:(g-2)e}, and it is a particularly clean probe since the same coupling combination enters both observables, $\kappa_e$ and $\Delta a_e$, as discussed in Sec.~\ref{subsec:flavourphysics}. The future prospects for $(g-2)_e$, represented by the blue bars, suppose an order of magnitude improvement in the precision of $\Delta a_e$. It corresponds to the limit of $\Delta a_e<5\cdot 10^{-14}$ which should be achieved by the time of the FCC-ee~\cite{Giudice:2012ms,DiLuzio:2024sps}.

The FCC-ee constraints are separated in two bars to show the impact of the $Z$-pole run (purple) and the additional run at $\SI{240}{\tera\electronvolt}$ (green). From the EW point of view, to extract the constraints we set the experimental value of each observable in Tab.~\ref{tab:ewpos} to its SM prediction~\cite{Breso-Pla:2021qoe} and rescale the experimental uncertainties using~\cite{Bernardi:2022hny, DeBlas:2019qco}. However, as shown in Fig.~\ref{fig:newke_vllpairs}, constraints on the models that contain a pair of vector-like leptons are dominated by the $Z$-pole run and the electroweak fit. Specifically, great precision at the FCC-ee puts strong constraints on the couplings of the new vector-like leptons to the SM leptons and the Higgs doublet, which also enter $\kappa_e$. The reason is that each of these couplings, $\lambda_\Psi$, with $\Psi=\{E,\Delta_1,\Delta_3,\Sigma,\Sigma_1\}$, defined in Eqs.~\eqref{eq:L1int}--\eqref{eq:L5int}, determines the $Z$-coupling modifier to the SM leptons, as shown in Eqs.~\eqref{eq:deltagZnuL}--\eqref{eq:deltagZeR} and Tab.~\ref{tab:treelevelmatching_novarphi}. Moreover, the remaining coupling between two vector-like leptons, necessary to enhance the electron Yukawa coupling, contributes to $O_{\phi D}$ at the one-loop level. This induces a shift in the $W$-boson mass, to which the electroweak fit demonstrates significant sensitivity. Consequently, the increased precision in Higgs coupling measurements does not achieve a competitive impact compared to the precision of the observables listed in Tab.~\ref{tab:ewpos}.

\paragraph{Models with a vector-like lepton and a scalar. } 

Before presenting the results, we comment on the form of the total $\chi^2$-function for models with one scalar and one VLL. In particular, the constraints from electroweak and Higgs physics are not enough to put theoretically reasonable bounds on the coupling between two heavy states, $\lambda_{\Phi\Psi}$. It is therefore necessary to impose a theoretical constraint from the quantum stability of these simplified models. Indeed, for $\lambda_{\Phi\Psi} \gtrsim 3$, the models with a NP scalar and a VLL develop a Landau pole below $10^3$ TeV~\cite{Paradisi:2022vqp}. Overall, the total $\chi^2$ reads
\begin{equation}
    \chi^2_{\rm tot} = \chi^2_{\rm EWPO} + \chi^2_{\rm Higgs} + 3.84\lround \frac{\lambda_{\Phi\Psi}}{3}\rround^2\,.
    \label{eq:chi2_total}
\end{equation}
Additionally, as was already mentioned for models involving only scalars, we set the couplings not entering the coupling modifier $\kappa_e$ to their best-fit values. 

\begin{figure}[t]
    \centering
    \includegraphics[width=0.7\linewidth]{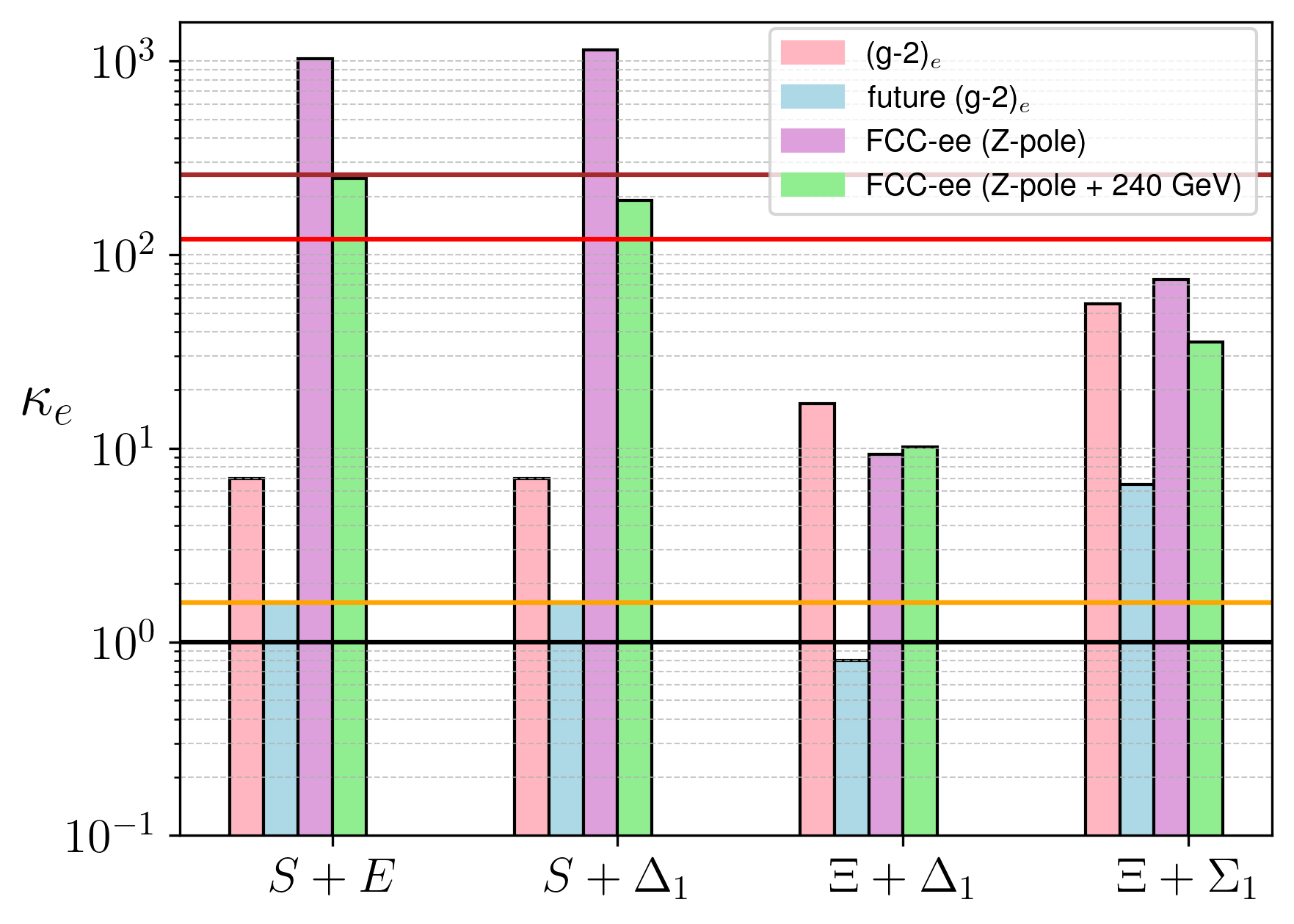}
    \caption{Final plot showing our results for the maximal allowed value for the coupling enhancement $\kappa_e$ with a 95\% CL for models with one VLL and one NP scalar. The auxiliary black line highlights the SM prediction $\kappa_e=1$. }
    \label{fig:newke_vllplusscalar}
\end{figure}

The results for models that involve a new physics scalar and a vector-like lepton are summarised in Fig.~\ref{fig:newke_vllplusscalar}. The models are defined in Sec.~\ref{subsec:models_vllplusscalar} and we recall that due to the constraints discussed in Sec.~\ref{subsec:flavourphysics}, we exclude Models SL5 and SL6 as possible candidates that produce large $\kappa_e$. Further, this class of models is categorized into two groups based on the scalar type: SL1 and SL2 feature a scalar singlet $S \sim (\mathbf{1},\mathbf{1})_0$, while models SL3 and SL4 contain a scalar $\Xi \sim (\mathbf{1},\mathbf{3})_0$ that is a triplet under $SU(2)$.

This categorisation is important to understand the results in Fig.~\ref{fig:newke_vllplusscalar}. Indeed, from Tab.~\ref{tab:treelevelmatching_novarphi}, one can notice that, out of the operators entering the electroweak fit ($\cO_{\phi e}$, $\cO_{\phi l}^{(1, \,3)}$, $\cO_{\phi WB}$, $\cO_{\phi D}$), none are generated by the scalar singlet $S$ at the tree-level, while the triplet $\Xi$ generates $\cO_{\phi D}$. However, both $S$ and $\Xi$ induce effects in $\cO_{\phi\Box}$ in the tree-level matching, making these models sensitive to constraints from modifications of the Higgs couplings. Thus, for Models SL1 and SL2, which do not affect the electroweak fit at tree level but do impact the Higgs fit, a distinction arises between the results obtained using only 
$Z$-pole projections (purple histograms) and those incorporating additional data from the $\SI{240}{\giga\electronvolt}$ center-of-mass energy run at FCC-ee (green histograms). Conversely, a strong distinction is absent for Models SL3 and SL4, where the constraints are already mostly dominated by the precision achieved during the $Z$-pole run.

The reduced impact of the electroweak fit renders Models SL1 and SL2 generally less constrained. For this reason, we add to Fig.~\ref{fig:newke_vllplusscalar} a further horizontal line (burgundy) associated with the upper limit of $y_e<260\, y_e^{\rm SM}$ found by ATLAS~\cite{ATLAS:2019old} to highlight that the current LHC program tests $\kappa_e$ at a similar level as the future FCC-ee runs (without a dedicated electron Yukawa measurement).

Besides the electroweak and Higgs observables, Fig.~\ref{fig:newke_vllplusscalar} demonstrates the power of $(g-2)_e$ measurements in probing the electron Yukawa coupling in this class of simplified models. This is because $\Delta a_e$ directly probes the combination of couplings that enter $\kappa_e$, including the coupling $\lambda_{\Phi\Psi}$, which is challenging to constrain using $\chi^2_{\rm tot}$ in Eq.~\eqref{eq:chi2_total}. Thus, in Models SL1-SL3, a precise assessment of $\Delta a_e$ at the level of $5\cdot 10^{-14}$ surpasses the precision in $\kappa_e$ achievable by a dedicated run at the FCC-ee~\cite{dEnterria:2021xij}. In Model SL4, a group-theoretic factor $\eta$ relating $\Delta_e$ and $\kappa_e$ in Eq.~\eqref{eq:ae-eta} is less than unity, leading to slightly larger $\kappa_e$ values inferred from future $(g-2)_e$ measurements.

%%%%%%%%%%%%%%%%%%%%%
\section{Conclusions}
\label{sec:conclusions}
%%%%%%%%%%%%%%%%%%%%%
A future FCC-ee collider offers the unique opportunity to measure the electron Yukawa coupling during a dedicated Higgs pole mass run. However, due to the timeline and cost of such a run, it is imperative to ask whether this run can provide new information in concrete model realizations where electron Yukawa couplings deviate significantly from the SM predictions.

We addressed this question by identifying simplified models that, within an EFT approach, lead to sizeable tree-level corrections to the electron Yukawa coupling. There exists one simplified model featuring just a single new field, namely an extra scalar doublet, and three classes of simplified models with two new states: two vector-like leptons, a vector-like lepton and a scalar singlet or triplet, and a scalar doublet together with a scalar singlet or triplet.

We considered HL-LHC projections for direct searches of new states, constraints from flavour physics, including projections from future experiments, and constraints arising from the $Z$ pole 
 and 240 GeV center-of-mass energy runs at the FCC-ee. We found that in all models with vector-like leptons, the $(g-2)_e$ provides competitive bounds, comparable to those from the FCC-ee. Specifically, for the models combining a scalar and a vector-like lepton, only one combination of new physics states is such that future $(g-2)_e$ measurements do not probe the model as effectively as the dedicated Higgs pole mass run at the FCC-ee could. In models with two vector-like leptons, the dedicated Higgs pole mass run can still probe unexplored parameter space. However, in most models, the parameter space remains small after considering the projections for the $(g-2)_e$ and other FCC-ee runs. In contrast, for models with only scalars, large modifications of the electron Yukawa coupling with respect to its SM value are possible, and a dedicated run at the Higgs pole mass can be considered an effective probe of the couplings in the scalar potential.

Overall, Figs.~\ref{fig:newke_vllpairs} and~\ref{fig:newke_vllplusscalar} highlight that future improvements in the measurement of $(g-2)_e$ could match the FCC-ee program in constraining the electron Yukawa coupling in concrete NP models. Hence, an important message from our study is that the correlated effects in EWPOs at the FCC-ee and $(g-2)_e$ may provide crucial insights into the nature of the underlying new physics responsible for modifying the electron Yukawa coupling.

Finally, we note that the models under consideration, being simplified, are not inherently compelling as they do not directly address any of the shortcomings of the SM. Moreover, they require a specific flavour structure that we do not attempt to explain here. However, it is worth emphasizing that many of the states considered here naturally emerge in UV-complete frameworks aimed at resolving SM problems and puzzles. For instance, additional scalar fields, such as an extra scalar doublet in two-Higgs-doublet extensions of the SM, arise in supersymmetric models~\cite{Haber:1984rc,Branco:2011iw}, in scenarios addressing the strong CP problem like DFSZ-type models~\cite{Zhitnitsky:1980tq,Dine:1981rt}, or in frameworks linked to dark matter~\cite{Cai:2013zga,Drozd:2014yla} and baryogenesis~\cite{Alanne:2016wtx,Han:2020ekm}. Similarly, vector-like leptons appear in strongly interacting theories related to the hierarchy problem~\cite{Arkani-Hamed:2002ikv,Agashe:2004rs,Hagedorn:2011un,Frigerio:2018uwx,Goertz:2021ndf} and the flavour puzzle~\cite{Bordone:2017bld,Alonso:2018bcg,Fuentes-Martin:2022xnb,Lizana:2024jby}. 

Notwithstanding this, the identified simplified models demonstrate the existence of scenarios that can uniquely be probed by a dedicated Higgs pole mass at the FCC-ee. 

% % % % % % % % % % % % 
\section*{Acknowledgments}
We thank Luca Di Luzio and Paride Paradisi for useful discussions on the $(g-2)_e$.  NS would like to thank Alejo Rossia and Alessandro Valenti for helpful discussions. This work received funding by the INFN Iniziative Specifiche APINE. RG acknowledges support from the Italian Ministry of University and Research (MUR) via the PRIN 2022 project n. 20225X52RA — MUS4GM2 funded by the European Union via the Next Generation EU package and is supported by the University of Padua under the 2023 STARS Grants@Unipd programme (Acronym and title of the project: HiggsPairs – Precise Theoretical Predictions for Higgs pair production at the LHC). BE would like to thank the Galileo Galilei Institute for Theoretical Physics for the hospitality and the INFN for partial support during the completion of this work.

% % % % % % % % % % % % 
\appendix
% % % % % % % % % % % % 
\section{Direct searches for VLLs}
\label{app:VLL_directsearches}

\begin{table}[t]
    \centering
    {\renewcommand{\arraystretch}{1.8}
    \begin{tabular}{|cc|ccccc|}
    \hline
     &    & $E$  &   $\Delta_1 = \begin{pmatrix} N' \\ E' \end{pmatrix}$ & $\Delta_3= \begin{pmatrix} E' \\ Y'\end{pmatrix}$ & $\Sigma = \begin{pmatrix} \Sigma^1 \\ \Sigma^2 \\ \Sigma^3\end{pmatrix}$ & $\Sigma_1 = \begin{pmatrix} \Sigma_1^1 \\ \Sigma_1^2 \\ \Sigma_1^3\end{pmatrix}$  \\
     \hline
     \parbox[t]{2mm}{\multirow{4}{*}{\rotatebox[origin=c]{90}{Electric Charges $\mathcal{Q}$}}}&1   &  -   &         -     &     -     &    $T_1=\frac{1}{\sqrt{2}}\lround \Sigma^1 - i \Sigma^2 \rround$       &    -      \\
     &0   &  -   &      $N'$     &     -     &    $T_0=\Sigma^3$       &    $T_0=\frac{1}{\sqrt{2}}\lround\Sigma_1^1 - i \Sigma_1^2\rround$      \\
     &-1  & $E$  &      $E'$     &   $E'$    &    $T_{-1}=\frac{1}{\sqrt{2}}\lround \Sigma^1 + i \Sigma^2 \rround$       &    $T_{-1}= \Sigma_1^3$       \\
     &-2  & -    &        -      &   $Y'$    &      -    &    $T_{-2}=\frac{1}{\sqrt{2}}\lround\Sigma_1^1 + i \Sigma_1^2\rround$       \\
     \hline
    \end{tabular}
    }
    \caption{Electric charges of the components of the VLL multiplets of Tab.~\ref{tab:vectorlikeleptons}.}
    \label{tab:VLLelectriccharges}
\end{table}

In this appendix, we summarise the derivation of the couplings between the $W^\pm$ and $Z$ bosons and the VLL components required for the study of the processes shown in Fig.~\ref{fig:VLLproduction_hadroncolliders}. The Lagrangians in  Sec.~\ref{sec:effectivefieldtheory&UVmodels} are written in the interaction basis, which presents non-diagonal mass matrices.  Such mass matrices receive contributions from the VLL mass terms, the SM Yukawa sector, and the Yukawa-like interactions between a VLL and a SM lepton. The rotations required to cast the mass matrices into diagonal form also affect the remaining interactions. Therefore, the couplings between the $W^\pm$ and $Z$ bosons can only be read off the original Lagrangian once a rotation has been performed. The necessary procedure is formally identical to the one detailed in Ref.~\cite{Erdelyi:2024sls},  with minor adaptions. Hence, we directly present our results. 

For models SL1--SL4, we need to consider one new VLL multiplet per time, with results in Sec.~\ref{app:vllLHC_individual}.  For models L1--L4, we instead consider the presence of two VLL multiplets per time in Sec.~\ref{app:vllLHC_pair}.

%%%%%%%%%%%%%%%%%%
\subsection{Single VLL multiplet}
\label{app:vllLHC_individual}
%%%%%%%%%%%%%%%%%%

\begin{table}[t]
    \centering
    \begin{subtable}[b]{0.45\textwidth}
    \centering
        \renewcommand{\arraystretch}{1.5}
        \begin{tabular}{|c|cccc|}
        \hline
        \multicolumn{5}{|c|}{$g_Z$} \\
        \hline
         $\mathcal{Q}_\Psi$ & E & $\Delta_1$ & $\Delta_3$ & $\Sigma_1$      \\
         \hline
         $0$                & -   & $1$        & -          & $2$ \\
         $-1$               & $0$ & $-1$       & $1$        & $0$ \\
         $-2$               & -   &  -         & $-1$       & $-2$ \\
         \hline
        \end{tabular}
        \caption{Values of $g_Z$.}
        \label{tab:gZ_singleVLL}
    \end{subtable}    
    \hfill
    \begin{subtable}[b]{0.45\textwidth}
    \renewcommand{\arraystretch}{1.5}
    \begin{tabular}{|cc|cccc|}
    \hline
        \multicolumn{6}{|c|}{$g_W$} \\
    \hline
    $\mathcal{Q}_{\Psi_1}$ & $\mathcal{Q}_{\Psi_1}$ & $E$ & $\Delta_1$ & $\Delta_3$ & $\Sigma_1$      \\
    \hline    
    -1                     &    0                   & -   &    1       &   -        &  $-\sqrt{2}$             \\     
    -2                     &    -1                  & -   &    -       &   1        &  $\sqrt{2}$             \\        
    \hline
    \end{tabular}
    \caption{Values of $g_W$.}
    \label{tab:gW_singleVLL}
    \end{subtable}
    \caption{Values for $g_W$ and $g_Z$ for models involving only one VLL per time. For compactness, the table provides the coupling for the VLL components according to their electric charge $\mathcal{Q}_\Psi$.}
\end{table}
Since lepton number violation constraints prevent the vector-like lepton $\Sigma\sim(\mathbf{1},\mathbf{3})_0$ from significantly contributing to the electron Yukawa coupling, the four VLL multiplets that we consider are the singlet $E$, the doublets $\Delta_1$ and $\Delta_3$, and the triplet $\Sigma_1$. Notation-wise, for the interactions with the Z boson, we write the Lagrangian as 
\begin{equation}
    \frac{\cos\theta_W}{g_L} \cL^r_Z = \bar{\Psi} \slashed{Z} \lsquare \frac{1}{2}g_Z - \mathcal{Q}_\Psi \sin^2\theta_W\rsquare \Psi\,,
\end{equation}
where $\mathcal{Q}_\Psi$ is the electric charge of the VLL $\Psi$. The different values of $\mathcal{Q}_\Psi$ are reported in Tab.~\ref{tab:VLLelectriccharges}, while the values of $g_Z$ are set in Tab.~\ref{tab:gZ_singleVLL}. Similarly, the interactions with the $W^\pm$ boson are written as
\begin{equation}
    \frac{\sqrt{2}}{g_L} \cL^r_W = g_W\bar{\Psi}_1 \slashed{W}^+ \Psi_2 + \hc\,,
\end{equation}
with $\Psi_1$ and $\Psi_2$ being two components of the same VLL multiplet. The possible values for $g_W$ are listed in Tab.~\ref{tab:gW_singleVLL}. The derived interactions result in the limits on the mass of new states shown in Tab.~\ref{tab:vll_massbounds}.

%%%%%%%%%%%%%%%%%%%%%%%%
\subsection{A pair of VLL multiplets} \label{app:vllLHC_pair}
%%%%%%%%%%%%%%%%%%%%%%%%

As soon as more than one VLL multiplet is introduced, interactions of the form $\psi_1\psi_2 W$ and $\psi_1 \psi_2 Z $, with $\psi_1$ and $\psi_2$ components of two different VLL multiplets are possible in the mass basis. Organising such components according to their electric charge, the interactions with the $Z$ boson can be generically written as 
\begin{equation}
    \frac{\cos\theta_W}{g_L} \cL^r_Z = \bar{\psi} \slashed{Z} \lsquare \frac{1}{2} G_Z- \mathcal{Q}_\psi \sin^2\theta_W \mathbb{1}\rsquare \psi \,,
    % \quad \lsquare G_Z \rsquare = \lsquare \mathbb{1} \rsquare \,. 
\end{equation}
with the non-diagonal matrix $G_Z$.

Though the rotations performed involve both vector-like leptons and SM leptons, the results reported below only display the VLL components for more compact expressions.
\begin{itemize}
    \item Model L1:
    \begin{align}
        \frac{\cos\theta_W}{g_L} \cL^r_Z  =& \begin{pmatrix} \bar{E}' & \bar{E} \end{pmatrix}\slashed{Z} \lsquare \frac{1}{2} \begin{pmatrix}
		-\frac{1}{2} & -\frac{1}{2}\\
		 -\frac{1}{2}  & -\frac{1}{2} 
	\end{pmatrix} - \lround-1\rround s^2_W\, \mathbb{1}\rsquare  \begin{pmatrix}
	    E' \\ E
	\end{pmatrix} + \nonumber\\
	& +  \bar{N}' \slashed{Z} \lround \frac{1}{2}- 0 s^2_W\rround  N'  + O\lround \frac{v}{M_\Psi} \rround \,.
    \end{align}
    \item Model L2: 
    \begin{align}
	\frac{\cos\theta_W}{g_L} \cL^r_Z =& \begin{pmatrix}\bar{E}' & \bar{E} \end{pmatrix}\slashed{Z} \lsquare \frac{1}{2} \begin{pmatrix}
		\frac{1}{2} & \frac{1}{2}\\
		\frac{1}{2}  & \frac{1}{2} 
	\end{pmatrix} - \lround-1\rround s^2_W\rsquare \begin{pmatrix}
		 E' \\ E
	\end{pmatrix} + \nonumber\\
	& + \bar{Y}' \slashed{Z} \lround -\frac{1}{2} - (-2)s^2_W \rround Y'+ O\lround \frac{v}{M_\Psi} \rround  \,.
\end{align}
    \item Model L3:
    \begin{align}
	\frac{\cos\theta_W}{g_L} \cL^r_Z =& \begin{pmatrix} \bar{E}' & \bar{T}_{-1}\end{pmatrix}\slashed{Z} \lsquare \frac{1}{2} \begin{pmatrix}
		\frac{1}{2} & \frac{1}{2} \\
		\frac{1}{2}  & \frac{1}{2} 
	\end{pmatrix} - \lround-1\rround s^2_W \mathbb{1}\rsquare \begin{pmatrix}
		E' \\ T_{-1}
	\end{pmatrix} + \nonumber\\
	& + \bar{T}_{0} \slashed{Z} \lround \frac{1}{2} \,  2  - 0 s^2_W\rround T_{0} + \nonumber\\
	&+ \begin{pmatrix} \bar{Y}' & \bar{T}_{-2} \end{pmatrix} \lsquare \frac{1}{2} \begin{pmatrix}
		-\frac{3}{2} & \frac{1}{2} \\
		\frac{1}{2}  & -\frac{3}{2}
	\end{pmatrix} - (-2)s^2_W \rsquare \begin{pmatrix} Y' \\ T_{-2} \end{pmatrix} + O\lround \frac{v}{M_\Psi} \rround\,.
\end{align}
    \item Model L4:
    \begin{align}
	\frac{\cos\theta_W}{g_L} \cL^r_Z =& \begin{pmatrix} \bar{E}' & \bar{T}_{-1}\end{pmatrix}\slashed{Z} \lsquare \frac{1}{2} \begin{pmatrix}
		 -\frac{1}{2} & -\frac{1}{2} \\
	   -\frac{1}{2}  & -\frac{1}{2} 
	\end{pmatrix} - \lround-1\rround s^2_W \mathbb{1}\rsquare \begin{pmatrix}
		E' \\ T_{-1}
	\end{pmatrix} + \nonumber\\
	& + \begin{pmatrix}  \bar{N}' & \bar{T}_{0} \end{pmatrix} \slashed{Z} \lsquare \frac{1}{2}\begin{pmatrix}
		\frac{3}{2}   &   -\frac{1}{2}   \\
		-\frac{1}{2}   &   \frac{3}{2} \\
	\end{pmatrix} - 0 s^2_W \rsquare \begin{pmatrix}  N' \\ T_{0}\end{pmatrix} + \nonumber\\
	& + \bar{T}_{-2} \, \slashed{Z} \lsquare -1 - (-2)s^2_W\rsquare T_{-2} + O\lround \frac{v}{M_\Psi} \rround \,.
\end{align}
\end{itemize}
For the interaction with the $W^\pm$ bosons, one can proceed similarly. The most general form of the interaction is provided by
\begin{equation}
    \frac{\sqrt{2}}{g_L} \cL^r_W = \bar{\psi}_1 \slashed{W} G_W \psi_2\,,  
\end{equation}
where $G_W$ is a dim$\lround\psi_1\rround\times$dim$\lround\psi_2 \rround$ -- dimensional matrix.\\
We list below the model-dependent results:
\begin{itemize}
    \item Model L1:
    \begin{align}
	\frac{\sqrt{2}}{g_L}\cL_W^r =& \begin{pmatrix}  \bar{E}' & \bar{E} \end{pmatrix} \slashed{W}^-\begin{pmatrix}
		\frac{1}{\sqrt{2}} \\
		\frac{1}{\sqrt{2}}
	\end{pmatrix} N' + O\lround \frac{v}{M_\Psi} \rround + \hc\,.
    \end{align}
    \item Model L2: 
    \begin{align}
	\frac{\sqrt{2}}{g_L}\cL_W^r = 
	 \begin{pmatrix} \bar{E}' & \bar{E} \end{pmatrix} \slashed{W}^+\begin{pmatrix}
		\frac{1}{\sqrt{2}} \\
		\frac{1}{\sqrt{2}}
	\end{pmatrix} Y'+ O\lround \frac{v}{M_\Psi} \rround + \hc\,.
    \end{align}
        \item Model L3:
    \begin{align}
        \frac{\sqrt{2}}{g_L}\cL_W^r =  & \begin{pmatrix}\bar{E}' & \bar{T}_{-1} \end{pmatrix} \slashed{W}^-\begin{pmatrix}
		1  \\
		-1
	\end{pmatrix} T_{0}+ \nonumber\\
	&+ \begin{pmatrix} \bar{E}' & \bar{T}_{-1} \end{pmatrix} \slashed{W}^+\begin{pmatrix}
		\frac{1}{2}+\frac{1}{\sqrt{2}} & \frac{1}{2}-\frac{1}{\sqrt{2}} \\
		\frac{1}{2}-\frac{1}{\sqrt{2}} & \frac{1}{2}+\frac{1}{\sqrt{2}}
	\end{pmatrix} \begin{pmatrix}Y' \\ T_{-2}\end{pmatrix} + O\lround \frac{v}{M_\Psi} \rround+\hc \,.
    \end{align}
    \item Model L4:
    \begin{align}
        \frac{\sqrt{2}}{g_L}\cL_W^r =  & \begin{pmatrix} \bar{E}' & \bar{T}_{-1} \end{pmatrix} \slashed{W}^-\begin{pmatrix}
		\frac{1}{2}-\frac{1}{\sqrt{2}} & \frac{1}{2}+\frac{1}{\sqrt{2}}  \\
		\frac{1}{2}+\frac{1}{\sqrt{2}} & \frac{1}{2}-\frac{1}{\sqrt{2}}
	\end{pmatrix}\begin{pmatrix} N' \\ T_{0}\end{pmatrix} + \nonumber\\
	&+ \begin{pmatrix}\bar{E}' & \bar{T}_{-1} \end{pmatrix} \slashed{W}^+\begin{pmatrix}
		-1 \\
		1
	\end{pmatrix} T_{-2} + O\lround \frac{v}{M_\Psi} \rround + O\lround \frac{v}{M_\Psi} \rround +  \hc\,.
    \end{align}
\end{itemize}
The bounds on the masses for each of the models are summarised in Tab.~\ref{tab:vll_massbounds}.

 \bibliographystyle{utphys.bst}
\bibliography{bibliography}

\end{document}